\documentclass[a4paper, 12pt]{article}

\usepackage[top=1in, bottom=1in, left=1in, right=1in]{geometry}
\usepackage{graphicx}
\usepackage{amsthm}
\usepackage{amsmath}
\usepackage{amsfonts}
\usepackage{amssymb}
\usepackage{mathtools}
\usepackage{pdfpages}
\usepackage{hyperref}
\usepackage{natbib}
\usepackage{fancyhdr}
\usepackage[english]{babel}
\usepackage{txfonts}
\usepackage{gensymb} %for degrees C
\usepackage{caption}
\usepackage{listings,relsize} 
\usepackage{color}
\usepackage{chngcntr}
\usepackage[normalem]{ulem}
\usepackage{cancel}
\usepackage{lineno}
\usepackage{soul}
\usepackage{array}
\usepackage{varwidth}
\usepackage{longtable}
\usepackage{tabularx}
\usepackage{booktabs}
\usepackage{ragged2e}
\usepackage{setspace}

\newcommand{\thinvert}{\,\vert\,}
\newcommand{\fatvert}{\;\vert\;}
\newcommand{\argmin}{\operatornamewithlimits{argmin}}

\newcommand{\jfdel} {\bgroup\markoverwith{\textcolor{red}{\rule[0.5ex]{2pt}{1.5pt}}}\ULon}

\newcolumntype{D}{m{20mm}}
\newcolumntype{L}{>{\raggedright\arraybackslash}m{33mm}}
\newcolumntype{C}{>{\centering\arraybackslash}m{9mm}}
\newcolumntype{R}{>{\raggedleft\arraybackslash}m{15mm}}
\newcolumntype{S}{>{\raggedleft\arraybackslash}m{12mm}}

\begin{document}

\singlespacing
%\linenumbers

\title{\normalsize{\textbf{Locally Adaptive Smoothing with Markov Random Fields and Shrinkage Priors}}}
\author{\small{\textbf{James R. Faulkner}} \\ 
\small{\textit{Quantitative Ecology and Resource Management}}\\ \small{\textit{University of Washington, Seattle, Washington, U.S.A.}} \and \small{\textbf{Vladimir N. Minin}} \\
\small{\textit{Departments of Statistics and Biology}}\\ \small{\textit{University of Washington, Seattle, Washington, U.S.A.}}}
\date{}

\maketitle

\begin{abstract}

We present a locally adaptive nonparametric curve fitting method that operates within a fully Bayesian framework. 
This method uses shrinkage priors to induce sparsity in order-$k$ differences in the latent trend function, providing a combination of local adaptation and global control. 
Using a scale mixture of normals representation of shrinkage priors, we make explicit connections between our method and 
$k$th order Gaussian Markov random field smoothing.
We call the resulting processes shrinkage prior Markov random fields (SPMRFs). 
We use Hamiltonian Monte Carlo to approximate the posterior distribution of model parameters because this method provides superior performance in the presence of the high dimensionality and strong parameter correlations exhibited by our models. 
We compare the performance of three prior formulations using simulated data and find the horseshoe prior provides the best compromise between bias and precision. 
We apply SPMRF models to two benchmark data examples frequently used to test nonparametric methods. 
We find that this method is flexible enough to accommodate a variety of data generating models and offers the adaptive properties and computational tractability to make it a useful addition to the Bayesian nonparametric toolbox.

\end{abstract}

\section{Introduction}

Nonparametric curve fitting methods find extensive use in many aspects of statistical modeling such as nonparametric regression, spatial statistics, and survival models, to name a few.  
Although these methods form a mature area of statistics, many computational and statistical challenges remain when such curve fitting needs to be incorporated into multi-level Bayesian models with complex data generating processes. 
This work is motivated by the need for a curve fitting method that could adapt to local changes in smoothness of a function, including abrupt changes or jumps, and would not be restricted by the nature of observations and/or their associated likelihood.  Our desired method should offer measures of uncertainty for use in inference, should be relatively simple to implement and computationally efficient. There are many methods available for nonparametric curve fitting, but few which meet all of these criteria. 
\par
Gaussian process (GP) regression \citep{neal1998, ras2006gp} is a popular Bayesian nonparametric approach for functional estimation that places a GP prior on the function of interest.  
The covariance function must be specified for the GP prior, and the isotropic covariance functions typically used are not locally adaptive.  Nonstationary covariance functions have been investigated to make GP regression locally adaptive \citep{brahim2004gp, paciorek2004nonstationary, paciorek2006spatial}.  
Any finite dimensional representation of GPs involves manipulations of, typically high dimensional, Gaussian vectors with mean vector and covariance matrix induced by the GP. 
Many GPs, including the ones with nonstationary covariance functions, suffer from high computational cost  imposed by manipulations (e.g., Cholesky factorization) of the dense covariance matrix  in the finite dimensional representation. 
\par
Sparsity can be imposed in the precision matrix (inverse covariance matrix)  by constraining a finite dimensional representation of a GP to be a Gaussian Markov random field (GMRF), and then computational methods for sparse matrices can be employed to speed computations \citep{rue2001fast, rue2005gaussian}.  Fitting smooth functions with GMRFs has been practiced widely.  These methods use difference equations as approximations to continuous function derivatives to induce smoothing, and have a direct relationship to smoothing splines \citep{speckman2003fully}.  GMRFs have also been used to develop Bayesian adaptive smoothing splines \citep{lang2002function, yue2012priors, yue2014bayesian}.  A similar approach is the nested GP \citep{zhu2013locally}, which puts a GP prior on the order-\textit{k} function derivative, which is in turn centered on another GP. This approach has good adaptive properties but has not been developed for non-Gaussian data.  
\par
Differencing has commonly been used as an approach to smoothing and trend estimation in time series analysis, signal processing, and spatial statistics.  Its origins go back at least to \cite{whittaker1922new}, who suggested a need for a trade off between fidelity to the data and smoothness of the estimated function. This idea is the basis of some frequentist curve-fitting methods based on penalized least squares, such as the smoothing spline \citep{reinsch1967smoothing, wahba1975smoothing} and the trend filter \citep{kim2009, tibshirani2014adaptive}.  These penalized least-squares methods are closely related to regularization methods for high-dimensional regression such as ridge regression \citep{hoerl1970ridge} and the lasso \citep{tibshirani1996lasso} due to the form of the penalties imposed.  
\par
Bayesian versions of methods like the lasso \citep{park2008bayesian} utilize shrinkage priors in place of penalties. 
Therefore, it is interesting to investigate how these shrinkage priors \citep{polson2010shrink,griffin2013some, dirlap2015} perform when applied to differencing-based time series smoothing. 
Although shrinkage priors have been used explicitly in the Bayesian nonparametric regression setting for regularization of wavelet coefficients \citep{abramovich1998wavelet, johnstone2005empirical, remenyi2015wavelet} and for shrinkage of order-$k$ differences of basis spline coefficients in adaptive Bayesian P-splines \citep{scheipl2009}, a Bayesian version of the trend filter and Markov random field (MRF) smoothing with shrinkage priors has not been thoroughly investigated.
To our knowledge, only \citet{roualdes2015bayesian}, independently from our work, looked at Laplace prior-based Bayesian version of the trend filter in the context of a normal response model. 
In this paper, we conduct a thorough investigation of smoothing with shrinkage priors applied to MRFs for Gaussian and non-Gaussian data.  
We call the resulting models shrinkage prior Markov random fields (SPMRFs).
\par
We borrow the idea of shrinkage priors from the sparse regression setting and apply it to the problem of function estimation. 
We take the perspective that nonparametric curve fitting is essentially a regularization problem where estimation of an unknown function can be achieved by inducing sparsity in its order-\textit{k} derivatives.  
We propose a few fully Bayesian variations of the trend filter \citep{kim2009, tibshirani2014adaptive} which utilize shrinkage priors on the $k$th-order differences in values of the unknown target function. 
The shrinkage imposed by the priors induces a locally adaptive smoothing of the trend. 
The fully Bayesian implementation allows representation of parameter uncertainty through posterior distributions and eliminates the need to specify a single global smoothing parameter by placing a prior distribution on the smoothing parameter, although complete automation is not possible so we offer ways to parameterize the global smoothing prior. 
In Section \ref{methodSection} we provide a derivation of the models starting from penalized frequentist methods and we show the relationship to GMRF models.  
In Section \ref{methodSection} we also describe our method of sampling from the posterior distribution of the parameters using Hamiltonian Monte Carlo (HMC), which is efficient and straight forward to implement.  
In Section \ref{simSection} we use simulations to investigate performance properties of the SPMRF models under two different prior formulations and we compare results to those for a GMRF with constant precision. 
We show that the choice of shrinkage prior will affect the smoothness and local adaptive properties.  
In Section \ref{dataSection} we apply the method to two example data sets which are well known in the nonparametric regression setting.

\section{Methods}
\label{methodSection}

\subsection{Preliminaries}
\label{methodPrelim}

We start by reviewing a locally adaptive penalized least squares approach to nonparametric regression known as the trend filter \citep{kim2009, tibshirani2011gen, tibshirani2014adaptive} and use that as a basis to motivate a general Bayesian approach that utilizes shrinkage priors in place of roughness penalties.  
We first consider the standard nonparametric regression problem to estimate the unknown function $f$.  
We let $\boldsymbol{\theta}$ represent a vector of values of $f$ on a discrete uniform grid $t \in \{1,2,\dots, n\}$, and we assume $\boldsymbol{y} = \boldsymbol{\theta} + \boldsymbol{\epsilon},$  where $\boldsymbol{\epsilon} {\sim} \text{N}(\textbf{0}, \textbf{I}\sigma^2)$, and $\boldsymbol{y}$ and $\boldsymbol{\epsilon}$ are vectors of length $n$.  
Here all vectors are column vectors.  Following \cite{tibshirani2014adaptive} with slight modification, the least squares estimator of the $k$th order trend filtering estimate $\boldsymbol{\hat{\theta}}$ is 
\begin{equation}
 \boldsymbol{\hat{\theta}} = \argmin_{\boldsymbol{\theta}} \; \, \lVert \textbf{y}-\boldsymbol{\theta}\rVert^2_2 + \lambda  \lVert \textbf{D}^{(k)}\boldsymbol{\theta}\rVert_1 \, ,  \label{tfilter}
\end{equation}
\noindent where $ \lVert \, \cdot \, \rVert_q$ represents the $L_q$ vector norm, and $\textbf{D}^{(k)}$ is an $(n-k) \times n$ forward difference operator matrix of order $k$, such that the $i$th element of the vector  $\Delta^k\boldsymbol{\theta} = \textbf{D}^{(k)}\boldsymbol{\theta}$ is the forward difference $\Delta^k \theta_i = (-1)^{k}\sum_{j=0}^{k}(-1)^j {{k}\choose {j}}\theta_{i + j}$.  
Note that   $\textbf{D}^{(k)}$ has recursive properties such that $\textbf{D}_{n}^{(k)} = \textbf{D}_{n-k+1}^{(1)}\textbf{D}_{n}^{(k-1)} $, where $\textbf{D}_{m}^{(h)}$ has dimensions $(m-h) \times m$.   
The objective function in equation (\ref{tfilter}) balances the trade-off between minimizing the squared deviations from the data (the first term in the sum on the right) with minimizing the discretized roughness penalty of the function $f$ (the second term in the sum on the right).  
The smoothing parameter $\lambda \ge 0$ controls the relative influence of the roughness penalty.  Setting $\lambda$ to 0 we get least squares estimation. 
As $\lambda$ gets large, the roughness penalty dominates, resulting in a function with \textit{k}-th order differences approaching 0 for all $t$.  
The trend filter produces a piecewise polynomial function of $t_1,\dots,t_n$ with degree $k-1$ as an estimator of the unknown function $f$.  
Increasing the order of the difference operator will enforce a smoother function.   
\par
The $L_1$ penalty in equation (\ref{tfilter}) results in the trend filter having locally adaptive smoothing properties.  \cite{tibshirani2014adaptive} shows that the trend filter is very similar in form and performance to smoothing splines and locally adaptive regression splines, but the trend filter has a finer level of local adaptivity than smoothing splines.  A main difference between the trend filter and smoothing splines is that the latter uses a squared $L_2$ penalty, which is the same penalty used in ridge regression \citep{hoerl1970ridge}.  Note that the $L_1$ penalty used by the trend filter is also used by the lasso regression \citep{tibshirani1996lasso}, and the trend filter is a form of generalized lasso \citep{tibshirani2011gen, tibshirani2014adaptive}.  
In the linear regression setting with regression coefficients $\beta_j$s, the $L_1$ and $L_2$ penalties can be represented by the generalized ridge penalty $\lambda\sum_{j}^{}|\beta_j|^q $ \citep{Frank1993}, where $q=2$ corresponds to the ridge regression penalty, $q=1$ to the lasso penalty, and sending $q$ to zero results in all subsets selection regression \citep{Tibshirani2011}.   
Based on what we know about lasso regression, subset selection regression, and ridge regression, we expect a penalty closer to subset selection to do better for fitting functions with a small number of large jumps, 
a trend filter penalty ($L_1$) to do better for fitting functions with small to moderate deviations from polynomials of degree $k-1$, and a smoothing spline (squared $L_2$) penalty to do better for smooth polynomial-like functions with no jumps.  
This distinction will become important later when we assess the performance of different Bayesian formulations of the trend filter. 
\par
One can translate the penalized least squares formulation in equation (\ref{tfilter}) into either a penalized likelihood formulation or a Bayesian formulation.  Penalized least squares can be interpreted as minimizing the penalized negative log-likelihood $-l_p(\boldsymbol{\theta}\mid\boldsymbol{y}) = -l(\boldsymbol{\theta}\mid \boldsymbol{y}) + p(\boldsymbol{\theta}\mid {\lambda} ),$ where $l(\boldsymbol{\theta}\mid \boldsymbol{y})$ is the unpenalized log-likelihood and $p(\boldsymbol{\theta}\mid {\lambda} )$ is the penalty.  It follows that maximization of the penalized log-likelihood is directly comparable to finding the mode of the log-posterior in the Bayesian formulation, where the penalty is represented as a prior. This implies independent Laplace (double-exponential) priors on the $\Delta^k\theta_j $, where $j = 1,\dots,n-k$, for the trend filter formulation in equation (\ref{tfilter}).  That is, $p(\Delta^k\theta_j \mid \lambda ) = \frac{\lambda}{2}\exp\left(-\lambda\thinvert\Delta^k\theta_j \mid \right). $ This is a well-known result that has been used in deriving a Bayesian form of the lasso \citep{tibshirani1996lasso, figueiredo2003adaptive, park2008bayesian}.  Note that putting independent priors on the $k$th order differences results in improper joint prior $p(\boldsymbol{\theta} \mid \lambda )$, which can be made proper by including a proper prior on the first $k$ $\theta$s.  
\par
The Laplace prior falls into a class of priors commonly known as shrinkage priors. An effective shrinkage prior has the ability to shrink noise to zero yet retain and accurately estimate signals \citep{polson2010shrink}.  These properties translate into a prior density function that has a combination of high mass near zero and heavy tails.  The high density near zero acts to shrink small values close to zero, while the heavy tails allow large signals to be maintained.  
A simple prior developed for subset selection in Bayesian setting is the spike-and-slab prior, which is a mixture distribution between a point mass at zero and a continuous distribution \citep{mitchell1988bayesian}.  This prior works well for model selection, but some drawbacks are that it forces small signals to be exactly zero, and computational issues can make it difficult to use \citep{polson2010shrink}.  There has been much interest in developing priors with continuous distributions (one group) that retain variable selection properties of the spike-and-slab (two-group) yet do so by introducing sparsity through shrinkage \citep{polson2010shrink}.  This approach allows all of the coefficients to be nonzero, but most are small and only some are large.  Many such shrinkage priors have been proposed, including the normal-gamma \citep{griffin2010inference}, generalized double-Pareto \citep{armagan2013generalized}, horseshoe \citep{carvalho2010horseshoe}, horseshoe+ \citep{bhadra2015hplus}, and Dirichlet-Laplace \citep{dirlap2015}.  The Laplace prior lies somewhere between the normal prior and the spike-and-slab in its shrinkage abilities, yet most shrinkage priors of current research interest have sparsity inducing properties closer to those of the spike-and-slab. 
Our main interest is in comparing the Laplace prior to other shrinkage priors in the context of nonparametric smoothing.

\subsection{Model Formulation}
\label{methodModel}

It is clear that shrinkage priors other than the lasso could represent different smoothing penalties and therefore could lead to more desirable smoothing properties.  
There is a large and growing number of shrinkage priors in the literature.  
It is not our goal to compare and characterize properties of Bayesian nonparametric function estimation under all of these priors.  
Instead, we wish to investigate a few well known shrinkage priors and demonstrate as proof of concept that adaptive functional estimation can be achieved with shrinkage priors.  
Further research can focus on improvements to these methods.  What follows is a general description of our modeling approach and the specific prior formulations that will be investigated through the remainder of the paper.
\par
We assume the $n$ observations $y_i$, where $i=1,\dots,n$, are independent and follow some distribution dependent on the unknown function values $\theta_i$ and possibly other parameters $\boldsymbol{\xi}$ at discrete points $t$. We further assume that the order-\textit{k} forward differences in the function parameters, $\Delta^k\theta_j$, where $j=1,\dots,n-k$, are independent and identically distributed conditional on a global scale parameter which is a function of the smoothing parameter $\lambda$.  These assumptions result in the following general hierarchical form:
\begin{equation}
\begin{split}
  y_i \fatvert \theta_i, \boldsymbol{\xi} \sim p(y_i \fatvert \theta_i, \boldsymbol{\xi}), 
 \quad  \Delta^k\theta_j \fatvert \lambda \sim p(\Delta^k\theta_j\fatvert\lambda), \quad
  \lambda \sim p(\lambda), \quad
  \boldsymbol{\xi}  \sim p(\boldsymbol{\xi}).  
\end{split}
\label{genform}
\end{equation} 
\par
One convenient trait of many shrinkage priors, including the Laplace, the logistic, and the \textit{t}-distribution, is that they can be represented as scale mixtures of normal distributions \citep{andrews1974scale, west1987scale,polson2010shrink}. 
The conditional form of scale mixture densities leads naturally to hierarchical representations.  
This can allow some otherwise intractable density functions to be represented hierarchically with standard distributions and can ease computation. 
To take advantage of this hierarchical structure, we restrict densities $p(\Delta^k\theta_j\fatvert\lambda)$ to be scale mixtures of normals, which allows us to induce a hierarchical form to our model formulation by introducing latent local scale parameters, $\tau_j$.  
Here the order-\textit{k} differences in the function parameters, $\Delta^k\theta_j$, are conditionally normally distributed with mean zero and variance $\tau_j^2$, and the $\tau_j$ are independent and identically distributed with a global scale parameter which is a function of the smoothing parameter $\lambda$.  The distribution statement for $\Delta^k\theta_j$ in Equation (\ref{genform}) can then be replaced with the following hierarchical representation: 
\begin{equation}
\begin{split}
 \quad  \Delta^k\theta_j \fatvert \tau_j \sim \text{N}(0, \tau_j^2), \quad
  \tau_j \fatvert \lambda \sim p(\tau_j \fatvert \lambda). \quad
\end{split}
\label{hierform}
\end{equation} 

\par
To complete the model specification, we place proper priors on $\theta_1, \dots, \theta_k$.  
This maintains propriety and can improve computational performance for some Markov chain Monte Carlo (MCMC) samplers. 
We start by setting $\theta_1 \sim \text{N}(\mu, \omega^2)$, where $\mu$ and $\omega$ can be constants or allowed to follow their own distributions.  Then for $k \geq 2$ and $h = 1, \dots, k-1$, we let $\Delta^h\theta_1\thinvert\alpha_h \sim \text{N}(0, \alpha_h^2)$ and $\alpha_h \thinvert \lambda \sim p(\alpha_h\thinvert \lambda)$, where $p(\alpha\thinvert\lambda) $ is the same form as $p(\tau \thinvert \lambda)$.  That is, we assume the order-$h$ differences are independent with scale parameters that follow the same distribution as the order-$k$ differences. For most situations, the order of $k$ will be less than 4, so issues of scale introduced by assuming the same distribution on the scale parameters for the lower and higher order differences will be minimal.  One could alternatively adjust the scale parameter of each $p(\alpha_h\thinvert\lambda)$ to impose smaller variance for lower order differences.  
 \par
For the remainder of the paper we investigate two specific forms of shrinkage priors: the Laplace and the horseshoe. We later compare the performance of these two priors to the case where the order-\textit{k} differences follow identical normal distributions.  The following provides specific descriptions of our shrinkage prior formulations.

\begin{figure}[t]
	\begin{center}
		\includegraphics[width=\textwidth]{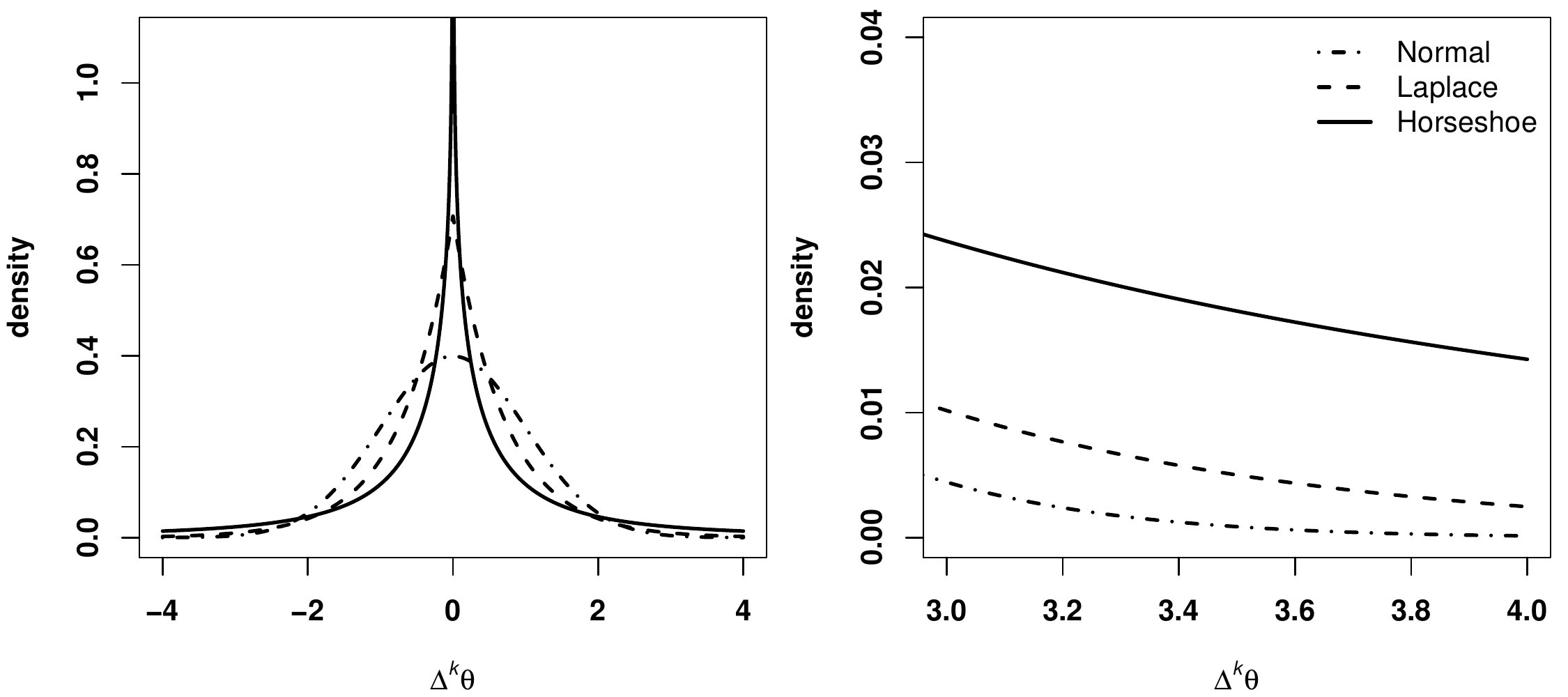}
	\end{center}
	\caption{Shapes of prior distributions (left) and associated tail behavior (right) for priors used for $p(\Delta^k\theta \thinvert \lambda) $.  \label{priorFig}  }
\end{figure}

\textit{Laplace}. As we showed previously, this prior arises naturally from an $L_1$ penalty, making it the default prior for Bayesian versions of the lasso \citep{park2008bayesian} and trend filter.  
The Laplace distribution is leptokurtic and features high mass near zero and exponential tails (Figure \ref{priorFig}). Various authors have investigated its shrinkage properties \citep{griffin2010inference, kyung2010penalized, armagan2013generalized}.  
We allow the order-\textit{k} differences $\Delta^k\theta_j$ to follow a Laplace distribution conditional on a global scale parameter $\gamma = 1/\lambda$, and we allow $\gamma$ to follow a half-Cauchy distribution with scale parameter $\zeta$.  
That is,
\begin{equation}
 \Delta^k\theta_j \fatvert \gamma \sim \text{Laplace}(\gamma), \quad \quad \quad
  \gamma \sim \text{C}^+(0, \zeta). 
\label{genDE}
\end{equation} 
The use of a half-Cauchy prior on $\gamma$ is a departure from \cite{park2008bayesian}, who make $\lambda^2$ follow a gamma distribution to induce conjugacy in the Bayesian lasso.  
We chose to use the half-Cauchy prior on $\gamma$ because its single parameter simplifies implementation, it has desirable properties as a prior on a scale parameter \citep{gelman2006prior, polson2012half}, and it allowed us to be consistent across methods (see horseshoe specification below).  
The hierarchical form of the Laplace prior arises when the mixing distribution on the square of the local scale parameter $\tau_j$ is an exponential distribution.  
Specifically, we specify $\tau_j^2 \fatvert \lambda \sim \text{Exp}(\lambda^2/2)$ and  $\Delta^k\theta_j \fatvert \tau_j \sim \text{N}(0, \tau_j^2)$ in the hierarchical representation.
\par
\textit{Horseshoe}. The horseshoe prior \citep{carvalho2010horseshoe} has an infinite spike in density at zero but also exhibits heavy tails (Figure \ref{priorFig}).  This combination results in excellent performance as a shrinkage prior \citep{polson2010shrink}, and gives the horseshoe shrinkage properties more similar to the spike-and-slab variable selection prior than those of the Laplace prior.  We allow the order-\textit{k} differences $\Delta^k\theta_j$ to follow a horseshoe distribution conditional on global scale parameter $\gamma=1/\lambda$, and allow $\gamma$ to follow a half-Cauchy distribution with scale parameter $\zeta$.  That is,  
\begin{equation}
 \Delta^k\theta_j \fatvert \gamma \sim \text{HS}(\gamma), \quad \quad \quad
  \gamma \sim \text{C}^+(0, \zeta). 
\label{genhorse}
\end{equation} The horseshoe density function does not exist in closed form, but we have derived an approximate closed-form solution using the known function bounds (see Appendix \ref{appendixA}), which could be useful for application in some settings. \cite{carvalho2010horseshoe} represent the horseshoe density hierarchically as a scale mixture of normals where the local scale parameters $\tau_j$ are distributed half-Cauchy.  In our hierarchical version, the latent scale parameter $\tau_j \fatvert \gamma \sim \text{C}^+(0, \gamma)$ and then conditional on $\tau_j$ the distribution on the order-\textit{k} differences is $\Delta^k\theta_j \fatvert \tau_j \sim \text{N}(0, \tau_j^2)$.
\par
The horseshoe prior arises when the mixing distribution on the local scale parameter $\tau_j$ is half-Cauchy, which is a special case of a half-\textit{t}-distribution where degrees of freedom (\textit{df}) equal 1.  
Setting $df > 1$ would result in a prior with lighter tails than the horseshoe, and setting $0 < df < 1$ would result in heavier tails.  
We tested half-\textit{t} formulations with $df$ between 1 and 5 in test scenarios, but did not find an appreciable difference in performance relative to the horseshoe.  
We also attempted to place a prior distribution on the $df$ parameter, but found the data to be insufficient to gain information in the posterior for $df$ in our test scenarios, so we did not pursue this further.       

\textit{Normal}.  The normal distribution arises as a prior on the order-$k$ differences when the penalty in the penalized likelihood formulation is a squared $L_2$ penalty.  The normal prior is also the form of prior used in Bayesian smoothing splines.  The normal is not considered a shrinkage prior and does not have the flexibility to allow locally adaptive smoothing behavior.  We use it for comparison to demonstrate the local adaptivity allowed by the shrinkage priors.  For our investigations, the distribution on the order-\textit{k} differences and associated scale parameter is:    
\begin{equation}
\Delta^k\theta_j \fatvert \gamma \sim \text{N}(0, \gamma^2), \quad \quad \quad
\gamma \sim \text{C}^+(0, \zeta). 
\label{gennorm}
\end{equation}

\subsection{Connections to Markov Random Fields}
\label{methodGMRF}

Here we briefly show the models represented by (\ref{genform}) can be expressed with GMRF priors for $\boldsymbol{\theta}$ conditional on the local scale parameters $\boldsymbol{\tau}$. 
It is instructive to start with the normal increments model \eqref{gennorm}, which belongs to a class of
time series models known as autoregressive models of order $k$.  
\cite{rue2005gaussian} call this model a $k$-th order random walk and show that it is a GMRF with respect to a $k$-th order chain graph --- a graph with nodes $\{1,2,\dots,n\}$, where the nodes $i \ne j$ are connected by an edge if and only if $|i - j| \le k$. 
Since the normal model \eqref{gennorm} does not fully specify the joint distribution of $\boldsymbol{\theta}$, it is an intrinsic (improper)
GMRF. 
We make it a proper GMRF by specifying a prior density of the first $k$ components of $\boldsymbol{\theta}$, $p(\theta_1,\dots, \theta_k)$.
The Markov property of the model manifests itself in the following factorization:
%\[
\begin{equation*}
p(\boldsymbol{\theta}) = p(\theta_1,\dots, \theta_k)p(\theta_{k+1}\mid \theta_1,\dots,\theta_k) \cdots p(\theta_n \mid \theta_{n-1}, \dots, \theta_{n-k}).
\end{equation*}
%\]
Equipped with initial distribution $p(\theta_1,\dots, \theta_k)$, models \eqref{genhorse} and \eqref{genDE} also admit this factorization, so they are $k$-th order Markov, albeit not Gaussian models. 
However, if we condition on the latent scale parameters $\boldsymbol{\tau}$, both the Laplace and horseshoe models become GMRFs, or 
more specifically $k$-th order normal random walks. One important feature of these random walks is that each step in the walk has its own precision.
To recap, under prior specifications \eqref{genhorse} and \eqref{genDE} $p(\boldsymbol{\theta}\mid \gamma) $ is a 
non-Gaussian Markov field, while $p(\boldsymbol{\theta} \mid \boldsymbol{\tau}, \gamma) = p(\boldsymbol{\theta} \mid \boldsymbol{\tau}) $ is a GMRF.
 
\par      
Our GMRF point of view is useful in at least three respects.
First, GMRFs with constant precision have been used for nonparametric smoothing in many settings (see \cite{rue2005gaussian} for examples).  GMRFs with nonconstant precision have been used much less frequently, but one important application is to the development of adaptive smoothing splines by allowing order-$k$ increments to have nonconstant variances \citep{lang2002function, yue2012priors}.  The approach of these authors is very similar to our own but differs in at least two important ways.  First, we specify the prior distribution on the latent local scale parameters $\tau_j$ with the resulting marginal distribution of $\Delta^k\theta_j$ in mind, such as the Laplace or horseshoe distributions which arise as scale mixtures of normals.  This allows a better understanding of the adaptive properties of the resulting marginal prior in advance of implementation. In contrast, \cite{lang2002function} and \cite{yue2012priors} appear to choose the distribution on local scale parameters based on conjugacy and do not consider the effect on the marginal distribution of $\Delta^k\theta_j$. Second, we allow the local scale parameters $\tau_j$ to be independent, whereas \cite{lang2002function} and \cite{yue2012priors} impose dependence among the scale (precision) parameters by forcing them to follow another GMRF.  Allowing the local scale parameters to be independent allows the model to be more flexible and able to adapt to jumps and sharp local features.  We should also note that \cite{rue2005gaussian} in section 4.3 show that the idea of scale mixtures of normal distributions can be used with GMRFs to generate order-\textit{k} differences which marginally follow a \textit{t}-distribution by introducing latent local scale parameters.  Although they do not pursue this further, we mention it because it bears similarity to our approach.       
\par
Second, viewing the SPMRF models as conditional GMRFs allows us to utilize some of the theoretical results and computational methods developed for GMRFs.
In particular, one can take advantage of more complex forms of precision matrices such as circulant or seasonal trend matrices (see \cite{rue2005gaussian} for examples).
One can also employ the computational methods developed for sparse matrices, which speed computation times \citep{rue2001fast, rue2005gaussian}.
We note that simple model formulations such as the $k$th-order random walk models can be coded with state-space formulations based on forward differences, which speed computation times by eliminating the operations on covariance matrices necessary with multivariate Gaussian formulations.
\par
A third advantage of connecting our models to GMRFs is that the GMRF representation allows us to connect our first-order Markov models to subordinated Brownian motion \citep{bochner1955, clark1973subordinated}, a type of L\'evy process recently studied in the context of scale mixture of normal distributions \citep{polson2012levy}. 
\cite{polson2012levy} use the theory of L\'evy processes to develop shrinkage priors and penalty functions. Let us briefly consider a simple example of subordinated Brownian motion.  
Let $W$ be a Weiner process, so that $W(t+s) - W(t) \sim \text{N}(0, s\sigma^2)$, and $W$ has independent increments.  
Let $T$ be a subordinator, which is a L\'evy process that is non-decreasing with probability 1, has independent increments, and is independent of $W$.  The subordinated process $Z$ results from observing $W$ at locations $T(t)$.  
That is, $Z(t) = W[T(t)]$.  The subordinator essentially generates a random set of irregular locations over which the Brownian motion is observed, which results in a new process.
In our hierarchical representation of Laplace and horseshoe priors \textit{for the first order differences}, we can define a subordinator process $T_j = \sum_{i=1}^j \tau_i^2$, so that
the GMRF $p(\boldsymbol{\theta} \mid \boldsymbol{\tau})$ can be thought of as a subordinated Brownian motion or as a realization of a Brownian motion with unit variance on the random latent irregular grid $T_1, \dots, T_n$.
The subordinated Brownian motion interpretation is not so straight forward when applied to higher-order increments, but we think this interpretation will be fruitful for extending our SPMRF models in the future. 
One example where this interpretation is useful is when observations occur on an irregularly spaced grid, which we explore in the following section.

\subsection{Extension to Irregular Grids}
\label{methodIrreg}
So far we have restricted our model formulation to the case where data are observed at equally-spaced locations. 
Here we generalize the model formulation to allow for data observed at locations with irregular spacing. 
This situation arises with continuous measurements over time, space, or some covariate, or when gaps are left by missing observations.
 
\par
For a GMRF with constant precision (normally distributed $k$th-order differences), we can use integrated Wiener processes to obtain the precision matrix (see \cite{rue2005gaussian} and \cite{lindgren2008second} for details).
However, properly accounting for irregular spacing in our models with Laplace or horseshoe $k$th-order differences is more difficult.  To use tools similar to those for integrated Wiener processes we would need to show that the processes built on Laplace and horseshoe increments maintain their distributional properties over any subinterval of a continuous measure.
\cite{polson2012levy} show that processes with Laplace or horseshoe first-order increments can be represented as subordinated Brownian motion.  
However, to meet the necessary condition of an infinitely divisible subordinator, the subordinator for the Laplace process needs to be on the precision scale and the subordinator for the horseshoe process needs to be on the log-variance scale.  
Both resulting processes are L\'evy process, which means they have independent and stationary increments, but the increments are no longer over the continuous measure of interest.  This makes representation of these processes over continuous time difficult and development of the necessary theory is out of the scope of this paper.  
\par
Absent theory to properly address this problem, we instead start with our hierarchical model formulations and assume that conditional on a set of local variance parameters $\boldsymbol{\tau}$, we can use methods based on integrated Wiener processes to obtain the precision matrices for the latent GMRFs.
This requires the assumption that local variances are constant within respective intervals between observations. 
Let $s_1 < s_2 < ... < s_n$ be a set of locations of observations, and let 
$\delta_j = s_{j+1} - s_{j}$ be the distance between adjacent locations.  
We assume we have a discretely observed continuous process and denote by $\theta(s_j)$ the value of the process at location $s_j$.  
For the first-order model and some interval $[s_j, s_{j+1}]$, we assume that conditional on local variance $\tau_j$, $\theta(s)$ follows a Wiener process where $\theta(s_j + h) - \theta(s_j) \mid \tau_j \sim \text{N}(0, h\tau_j^2) $ for all $0 \leq h \leq \delta_j $. If we let $\Delta\theta_j = \theta(s_{j+1}) - \theta(s_j)$, the resulting variance of $\Delta\theta_j$ is 
$$\text{Var}(\Delta\theta_j) = \delta_j\tau^2_j. $$ 
Note that the resulting marginal distribution of $\theta(s_j + h) - \theta(s_j)$ after integrating over $\tau_j$ is therefore assumed to be Laplace or horseshoe for all $h$, with the form of the marginal distribution dependent on the distribution of $\tau_j$. We know this cannot be true in general given the properties of these distributions, but we assume it approximately holds for $h \leq \delta$.
\par
The situation becomes more complex for higher order models.  We restrict our investigations to the second-order model and follow the methods of \cite{lindgren2008second}, who use a Galerkin approximation to the stochastic differential equation representing the continuous process. 
The resulting formula for a second-order increment becomes
$$ \Delta^2\theta_{j} = \theta(s_{j+2}) - \left( 1+ \frac{\delta_{j+1}}{\delta_j} \right) \theta(s_{j+1}) + \frac{\delta_{j+1}}{\delta_j}\theta(s_{j}), $$ 
and the variance of a second-order increment conditional on $\tau_j$ is
$$ \text{Var}(\Delta^2\theta_j) = \frac{\delta_{j+1}^2(\delta_j+\delta_{j+1})}{2}\tau^2_j. $$  This adjustment of the variance results in good consistency properties for GMRFs with constant precision \citep{lindgren2008second}, so should also perform well over intervals with locally constant precision.  We show in Appendices \ref{appendixA} and \ref{appendixB} that integrating over the local scale parameter $\tau_j$ maintains the distance correction as a multiplicative factor on the scale of the resulting marginal distribution.  We also provide a data example involving a continuous covariate in Appendix \ref{appendixC} where we apply the methods above for irregular grids.

\subsection{Posterior Computation}
\label{methodMCMC}
Since we have two general model formulations, marginal and hierarchical, we could use MCMC to approximate the posterior distribution of heights of our piecewise step functions, $\boldsymbol{\theta}$, by working with either one of the two corresponding posterior distributions. 
The first one corresponds to the marginal model formulation: 
\begin{equation}
p(\boldsymbol{\theta}, \gamma, \boldsymbol{\xi} \mid \mathbf{y}) \propto \prod_{i=1}^n p(y_i \mid \theta_i, \boldsymbol{\xi})p(\boldsymbol{\theta} \mid \gamma) p(\boldsymbol{\xi})p(\gamma),
\label{marginal-posterior}
\end{equation}
where $p(\boldsymbol{\theta} \mid \gamma)$ is a Markov field induced by the normal, Laplace, or horseshoe densities,  and p($\gamma$) is a half-Cauchy density.
Note that a closed-form approximation to the density function for the horsehoe prior (see Appendix \ref{appendixA}) is needed for the marginal formulation using the horseshoe.
The second posterior corresponds to the hierarchical model with latent scale parameters $\boldsymbol{\tau}$:
\begin{equation}
p(\boldsymbol{\theta}, \boldsymbol{\tau}, \gamma, \boldsymbol{\xi} \mid \mathbf{y}) \propto \prod_{i=1}^n p(y_i \mid \theta_i, \boldsymbol{\xi})p(\boldsymbol{\theta} \mid \boldsymbol{\tau}) \prod_{j=1}^{n-k} p(\tau_j \mid \gamma) p(\boldsymbol{\xi})p(\gamma),
\label{hierarchical-posterior}
\end{equation}
where $p(\boldsymbol{\theta} \mid \boldsymbol{\tau})$ is a GMRF and the choice of $p(\tau_j \mid \gamma)$ makes the marginal prior specification for $\boldsymbol{\theta}$ correspond either to a Laplace or to a horseshoe Markov random field. 
Notice that the unconditional GMRF (normal prior) has only the marginal specification.
\par  
Both of the above model classes are highly parameterized with dependencies among parameters induced by differencing and the model hierarchy.  
It is well known that high-dimensional, hierarchical models with strong correlations among parameters can create challenges for standard MCMC samplers, such as component-wise random walk Metropolis or Gibbs updates.  
When faced with these challenges, random walk behavior can result in inefficient exploration of the parameter space, which can lead to poor mixing and prohibitively long convergence times. 
Many approaches have been proposed to deal with these issues, including block updating \citep{knorr2002block}, elliptical slice sampling \citep{murray2010elliptical, murray2010slice}, the Metropolis adjusted Langevin algorithm (MALA) \citep{roberts2002langevin}, and Hamiltonian Monte Carlo (HMC) \citep{duane1987hybrid,neal1993probabilistic, neal2011mcmc}.  
All of these approaches jointly update some or all of the parameters at each MCMC iteration, which usually improves mixing and speeds up convergence of MCMC.  
Among these methods, HMC offered the most practical choice due to its ability to handle a wide variety of models and its relative ease in implementation via readily availble software such as \texttt{stan} \citep{stan2016}.  
We used a modification of HMC proposed by \cite{hoffman2014no} which automatically adjusts HMC tuning parameters. 
We used the open source package \texttt{rstan} \citep{rstan-software:2015}, which provides a platform for fitting models using HMC in the \texttt{R} computing environment \citep{rmanual}. 
 \par 
Even with HMC, slow mixing can still arise with hierarchical models and heavy-tailed distributions due to the inability of a single set of HMC tuning parameter values to be effective across the entire model parameter space. 
Fortunately this problem can often be remedied by model reparameterizations that change the geometry of the sampled parameter space.  For hierarchical models, the non-centered parameterization methods described by  \cite{papa2003non, papaspiliopoulos2007general} and \cite{betancourt2015hmc} can be useful.  
Non-centered parameterizations break the dependencies among parameters by introducing deterministic transformations of the parameters.  
The MCMC algorithm then operates directly on the independent parameters.
\cite{betancourt2015hmc} discuss non-centered parameterizations in the context of HMC, and further examples of these and other reparameterization methods that target heavy-tailed distributions are provided in the documentation for \texttt{stan} \citep{stan-manual:2015}.  
\par
We note that after employing reparameterizations, HMC with stationary distribution equal to the hierarchical model posterior \eqref{hierarchical-posterior} had good convergence and mixing properties for each of our models and in nearly all of our numerical experiments.  
HMC that targeted the marginal model posterior \eqref{marginal-posterior} had fast run times and good mixing for the normal and Laplace formulations, but we could not effectively reparameterize the (approximate) marginal horseshoe distribution to remove the effects of its heavy tails, which resulted in severe mixing problems for the marginal horseshoe-based model. 
Therefore, in the rest of the manuscript we work with the hierarchical model posterior distribution \eqref{hierarchical-posterior} for all models.
\par
{For SPMRF and GMRF models, the computation time needed to evaluate the log-posterior and its gradient scales as $\mathcal{O}(n)$, where $n$ is the grid size.  
However, the hierarchical SPMRF models have approximately twice as many parameters as the GMRF or marginal SPMRF models.  
These hierarchical SPMRF methods are therefore slower than their GMRF counterparts.
Since the computational cost of evaluating the log-posterior is only one factor determining the MCMC speed, we compared run times of the SPMRF and GMRF models on simulated and real data (see Appendix \ref{appendixG}). 
Our results show that SPMRF models are slower than GMRFs, but not prohibitively so.   
\par
We developed an \texttt{R} package titled \texttt{spmrf} which allows for easy implementation of our models via a wrapper to the \texttt{rstan} tools.  
The package code is publicly available at \url{https://github.com/jrfaulkner/spmrf}.

\section{Simulation Study}
\label{simSection}
\subsection{Simulation Protocol}

We use simulations to investigate the performance of two SPMRF formulations using the Laplace and horseshoe shrinkage priors described in Section \ref{methodModel} and compare results to those using a normal distribution on the order-$k$ differences.  We refer to the shrinkage prior methods as adaptive due to the local scale parameters, and the method with normal prior as non-adaptive due to the use of a single scale parameter.  We constructed underlying trends with a variety of characteristics following approaches similar to those of other authors \citep{scheipl2009, yue2012priors, zhu2013locally}.  We investigated four different types of underlying trend (constant, piecewise constant, smooth function, and function with varying smoothness). The first row of Figure \ref{simresplots} shows examples of the trend functions, each illustrated with simulated normal observations centered at the function values over a regular grid.  We used three observation types for each trend type where the observations were conditionally independent given the trend function values $\theta_i$, where $i=1,\dots,n$.  The observation distributions investigated were 1) normal: $y_i\thinvert\theta_i \sim \text{N}(\theta_i, \sigma^2)$, where  $\sigma = 1.5$ or $\sigma=4.5$; 2) Poisson: $y_i\thinvert\theta_i \sim \text{Pois}(\exp(\theta_i))$; and 3) binomial:  $y_i\thinvert\theta_i \sim \text{Binom}(m, (1+\exp(-\theta_i))^{-1} )$, where $m=20$ for all scenarios.
 
Note that we constructed the function values for the scenarios with normally distributed observations so that each function would have approximately the same mean and variance, where the mean and variance were calculated across the function values realized at the discrete time points.  This allowed us to specify observation variances which resulted in the same signal-to-noise ratio for each function, where signal-to-noise ratio is defined as the standard deviation of function values divided by the standard deviation of observations.  The signal-to-noise ratios for our scenarios with normal observations were 6 for $\sigma=1.5$ and 2 for $\sigma=4.5$. 
We chose the mean sizes for the Poisson scenarios and sample sizes for the binomial scenarios so that the resulting signal-to-noise ratios would be similar to those for the normal scenarios with $\sigma=4.5$. 
These levels allowed us to assess the ability of the models to adapt to local features when the signal is not overwhelmed by noise.  
We describe the trend functions further in what follows.
\par
\textit{Constant}.  This scenario uses a constant mean across all points.  We use this scenario to investigate the ability of each method to find a straight horizontal line in the presence of noisy data.  The values used for the constant mean were 20 for normal and Poisson observations, and 0.5 for binomial observations.  

\textit{Piecewise constant}.  This type of function has been used by \cite{tibshirani2014adaptive} and others such as \cite{scheipl2009} and \cite{zhu2013locally}.  The horizontal trends combined with sharp breaks offer a difficult challenge for all methods.  For the scenarios with normal or Poisson observations, the function values were 25, 10, 35, and 15 with break points at $t \in \{20, 40, 60\}$. For the binomial observations the function values on the probability scale were 0.65, 0.25, 0.85, and 0.45 with the same break points as the other observation types.  

\textit{Smooth trend}.  We use this as an example to test the ability of the adaptive methods to handle a smoothly varying function.  We generated the function $\boldsymbol{f}$ as a GP with squared exponential covariance function.  That is, 
$
\boldsymbol{f} \sim \text{GP}(\mu, \boldsymbol{\Sigma}), 
   \Sigma_{i,j} = \sigma_f^2\exp\left[-(t_j - t_i)^2/(2\rho^2)  \right],
$
\noindent where $\Sigma_{i,j}$ is the covariance between points $i$ and $j$, $\sigma_f^2 > 0$ is the signal variance and $\rho > 0$ is the length scale.  
We set $\mu=10$, $\sigma_f^2=430$, and $\rho=10$ for the scenarios with normal or Poisson observations.  
For binomial observations, $\boldsymbol{f}$ was generated in logit space with $\mu=-0.5$, $\sigma_f^2=3$, and $\rho=10$ and then back-transformed to probability space.  
For all scenarios the function was generated with the same random number seed. 
\par
\textit{Varying smoothness}. This function with varying smoothness was initially presented by \cite{dimatteo2001bayesian} and later used by others,
including \cite{yue2012priors}.  We adapted the function to a uniform grid, $t \in [1, n]$, where $n=100$ in our case, resulting in the function $$ g(t) = \sin\left( \frac{4t}{n}-2\right) + 2\exp\left(-30\left(\frac{4t}{n}-2\right)^2 \right). $$   For the normal and Poisson observations we made the transformation $f(t) = 20 + 10g(t)$.  For binomial observations we used $f(t) = 1.25g(t)$ on the logit scale. 
\par
We generated 100 datasets for each combination of trend and observation type. 
This number of simulations was sufficient to identify meaningful differences between models without excessive computation time.  
Each dataset had 100 equally-spaced sample points over the interval $[1,100]$.  
For each dataset we fit models representing three different prior formulations for the order-$k$ differences, which were 1) normal, 2) Laplace, and 3) horseshoe.  
We used the hierarchical prior representations for these models given in Section \ref{methodModel}.  
We selected the degree of $k$-th order differences for each model based on knowledge of the shape of the underlying function.  
We fit first-order models for the constant and piecewise constant functions, and we fit second-order models for the smooth and varying smooth functions. 
For the scenarios with normal observations, we set $\sigma \sim \text{C}^+(0,5)$.  
In all cases, $\theta_1 \sim \text{N}(\mu, \omega^2) $, where $\mu$ is set to the sample mean and $\omega$ is two times the sample standard deviation of the observed data transformed to match the scale of $\theta$.  
We also set $\gamma \sim \text{C}^+(0, 0.01) $ for all models.
\par
We used HMC to approximate the posterior distributions.  
For each model we ran four independent chains with different randomly generated starting parameter values and initial burn-in of 500 iterations. 
For all scenarios except for normal observations with $\sigma=1.5$, each chain had 2,500 posterior draws post-burn-in that were thinned to keep every 5th draw. 
For scenarios with normal observations with $\sigma=1.5$, chains with 10,000 iterations post-burn-in were necessary, with additional thinning to every 20th draw.
In all cases, these settings resulted in 2,000 posterior draws retained per model. We found that these settings consistently resulted in good convergence properties, where convergence and mixing were assessed with a combination of trace plots, autocorrelation values, effective sample sizes, and potential scale reduction statistics \citep{gelman1992inf}.     
\par
We assessed the relative performance of each model using three different summary statistics.  We compared the posterior medians of the trend parameters ($\hat{\theta}_i$) to the true trend values ($\theta_i$) using the mean absolute deviation (MAD): 
\begin{equation}	
  \text{MAD} = \frac{1}{n}\sum_{i=1}^{n} \vert \hat{\theta}_i - \theta_i \vert .
\end{equation} 

\noindent We assessed the width of the 95\% Bayesian credible intervals (BCIs) using the mean credible interval width (MCIW): 
\begin{equation}
	\text{MCIW} = \frac{1}{n}\sum_{i=1}^{n}  \hat{\theta}_{97.5, i} - \hat{\theta}_{2.5, i},
\end{equation} 

\noindent where $\hat{\theta}_{97.5, i}$ and $\hat{\theta}_{2.5, i}$ are the 97.5\% and 2.5\% quantiles of the posterior distribution for $\theta_i$. 
We also computed the mean absolute sequential variation (MASV) of $ \boldsymbol{\hat{\theta} } $ as 
\begin{equation}
 \text{MASV} = \frac{1}{n-1} \sum_{i=1}^{n-1} \vert \hat{\theta}_{i+1} - \hat{\theta}_i \vert .
\end{equation}  
We compared the observed MASV to the true MASV (TMASV) in the underlying trend function, which is calculated by substituting true $\theta$'s into equation for MASV.

\begin{figure}[th!]
	\begin{center}
		\includegraphics[width=\textwidth]{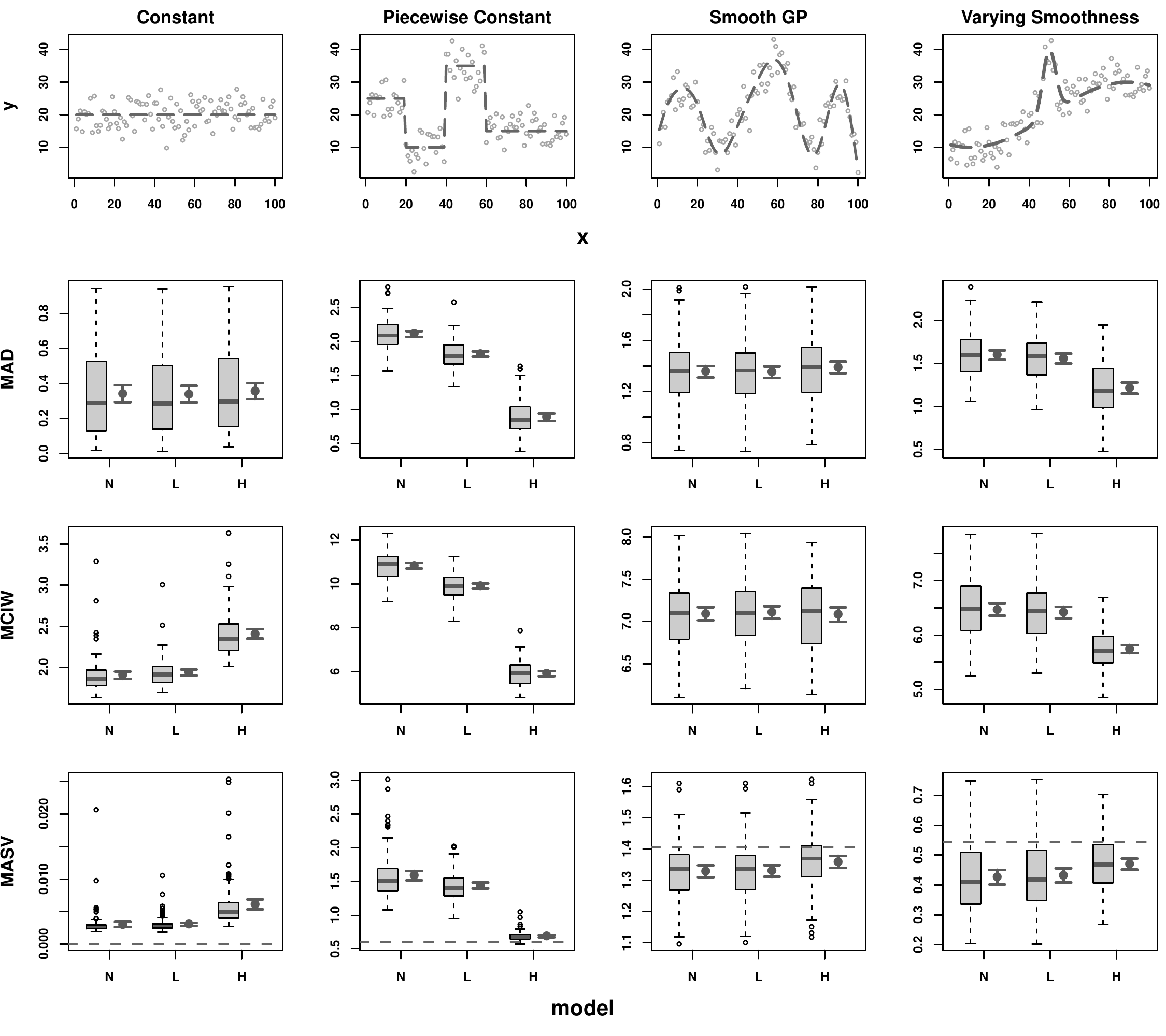}
	\end{center}
	\caption{Functions used in simulations and simulation results by model (N=normal, L=Laplace, H=horseshoe) and function type for normally distributed data with $\sigma$ = 4.5. Top row shows true functions (dashed lines) with example simulated data.  Remaining rows show mean absolute deviation (MAD), mean credible interval width (MCIW), and mean absolute sequential variation (MASV).  Horizontal dashed line in plots on bottom row is the true mean absolute sequential variation (TMASV). Shown for each model are standard boxplots of simulation results (left) and mean values with 95\% frequentist confidence intervals (right). \label{simresplots}  }
\end{figure}

\subsection{Simulation Results}

In the interest of space, we emphasize results for the scenarios with normally distributed observations with $\sigma=4.5$ here.  This level of observation variance was similar to that for Poisson and binomial observations and therefore offered results similar to those scenarios.   We follow these results with a brief summary of results for the other observation types, and we provide further summary of other results in Appendix \ref{appendixD}.

\begin{figure}[t] 
         \begin{center}
           \includegraphics[width=\textwidth]{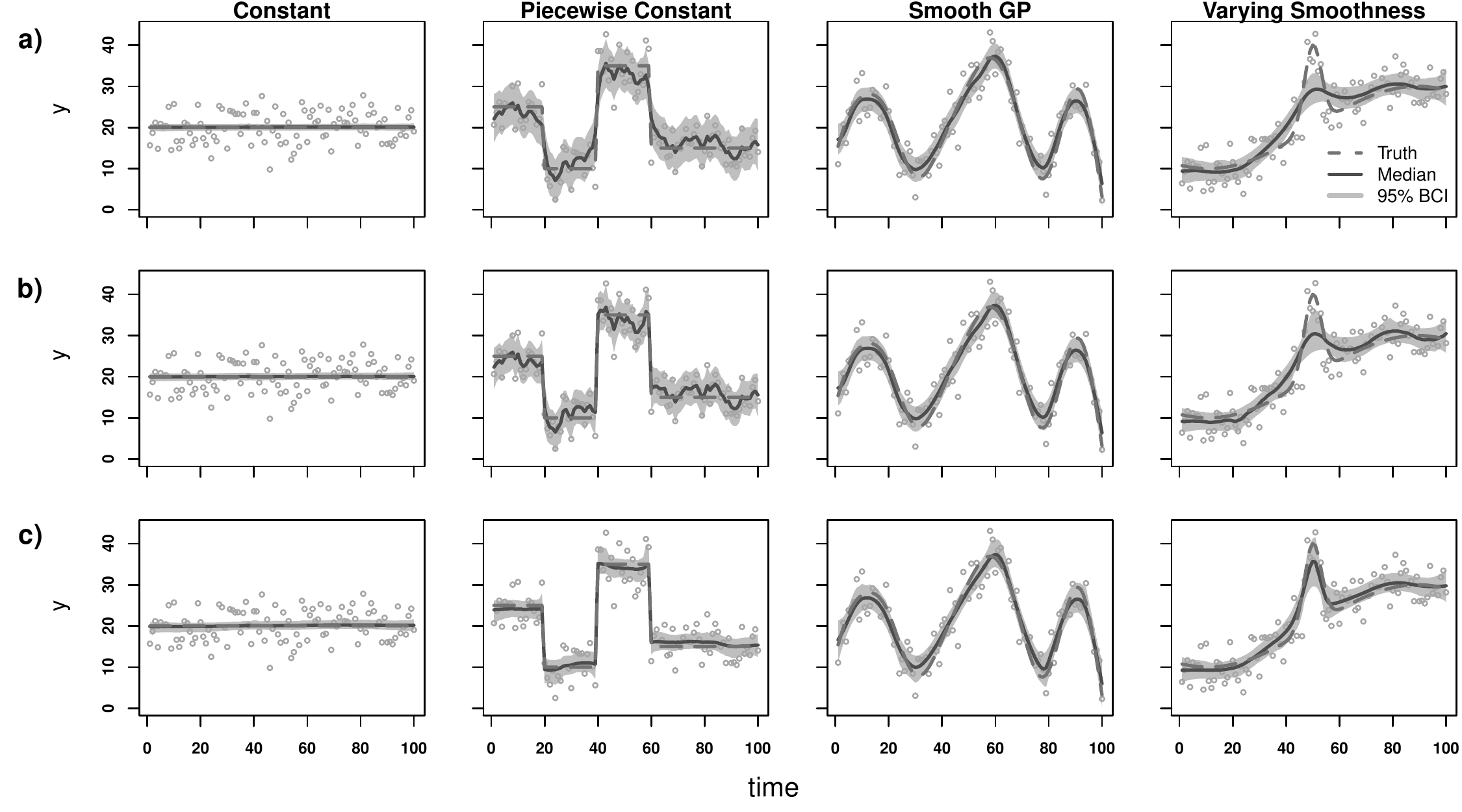}
         \end{center}
    \caption{Example fits for models using a) normal, b) Laplace, and c) horseshoe priors where observations are drawn from normal distributions with SD = 4.5.  Plots show true functions (dashed gray lines), posterior medians (solid dark gray lines), and associated 95\% Bayesian credible intervals (BCI; gray bands) for each $\theta$.  Values between observed locations are interpolated for plotting. }
   \label{exampfits}
\end{figure}

\textit{Constant}.  The three models performed similarly in terms of absolute value of all the metrics (Table \ref{normtab} and Figure \ref{simresplots}), but the Laplace and normal models were slightly better at fitting straight lines than the horseshoe.  
This is evidenced by the fact that the horseshoe had larger MCIW and larger MASV than the other methods.  
The first column of plots in Figure \ref{exampfits} provides a visual example of the extra variation exhibited by the horseshoe.

\begin{table}[t]
\caption{Mean values of performance measures across 100 simulations for normal observations ($\sigma=4.5$) for each model and trend function type. } \label{normtab}
\centering
\begin{tabular}{llrrrr}
    %\hline
    \toprule
 \textbf{Function} & \textbf{Model} & \textbf{MAD} & \textbf{MCIW} & \textbf{MASV} & \textbf{TMASV}  \\ 
    %\hline
    \midrule
Constant & Normal & 0.341 & 1.904 & 0.003 & 0.000  \\ 
   & Laplace & 0.339 & 1.937 & 0.003 & 0.000 \\ 
   & Horseshoe  & 0.356 & 2.406 & 0.006 & 0.000  \\ 
    %\hline
    \midrule
Piecewise Const. & Normal & 2.112 & 10.826 & 1.587 & 0.606\\ 
   & Laplace & 1.816 & 9.899 & 1.441 & 0.606 \\ 
   & Horseshoe & 0.886 & 5.919 & 0.689 & 0.606  \\ 
    %\hline
    \midrule
Smooth & Normal & 1.355 & 7.092 & 1.328 & 1.406 \\ 
   & Laplace & 1.352 & 7.106 & 1.329 & 1.406 \\ 
   & Horseshoe& 1.389 & 7.081 & 1.359 & 1.406  \\ 
    %\hline
    \midrule
Varying Smooth & Normal & 1.596 & 6.467 & 0.426 & 0.543 \\ 
   & Laplace &1.552 & 6.413 & 0.432 & 0.543 \\ 
   & Horseshoe &1.211 & 5.743 & 0.470 & 0.543 \\ 
    % \hline
    \bottomrule
\end{tabular}
\end{table}

\textit{Piecewise constant}.  The horseshoe model performed the best in all categories for this scenario and the normal model performed the worst (Table \ref{normtab} and Figure \ref{simresplots}).   The Laplace model was closer to the normal model in performance.  The horseshoe was flexible enough to account for the large function breaks yet still able to limit variation in the constant segments.  Example fits for the piecewise constant function are shown in the second column of plots in Figure \ref{exampfits}.

\textit{Smooth trend}.  The different models were all close in value of the performance metrics for the smooth trend scenario (Table \ref{normtab} and Figure \ref{simresplots}).  
The normal and Laplace models had smallest MAD, but the horseshoe had MSAV closer to the true MSAV.  
The fact that the values of the metrics were similar for all models suggests that not much performance is lost in fitting a smooth trend with the adaptive methods in comparison to non-adaptive. 

\textit{Varying Smoothness}.  Again the models all performed similarly in terms of absolute value of the metrics, but there was a clear ordering among models in relative performance (Table \ref{normtab} and Figure \ref{simresplots}). The horseshoe model performed the best relative to the other models on all metrics.  This function forces a compromise between having large enough local variance to capture the spike and small enough local variance to remain smooth through the rest of the function.  The horseshoe was more adaptive than the other two methods and therefore better able to meet the compromise. The plots in the last column of Figure \ref{exampfits} provide example fits for this function.

The results for the scenarios with normal observations with $\sigma=1.5$ and Poisson and binomial observations (see Appendix \ref{appendixD}) showed similar patterns to those with normal observations and $\sigma=4.5$.  For the constant function, the normal prior performed the best and the horseshoe prior the worst, although differences in terms of absolute values of the performance metrics were small. The relative differences were more pronounced with the scenarios with normal observations with $\sigma=1.5$. For the piecewise constant function, the horseshoe prior performed the best for all scenarios and the normal prior the worst.  All methods performed similarly for the smooth function, with the normal and Laplace generally performing a little better than the horseshoe.  For the function with varying smoothness, the horseshoe performed the best and the normal the worst for all scenarios.

\section{Data Examples}
\label{dataSection}

Here we provide two examples of fitting SPMRF models to real data.  
Each example uses a different probability distribution for the observations.  
The first example exhibits a change point, which makes it amenable to adaptive smoothing methods.  
The second example has a more uniformly smooth trend but also shows a period of rapid change, so represents a test for all methods. 
First we address the issue of setting the hyperparameter for the global smoothing parameter.

\subsection{Parameterizing the Global Smoothing Prior}
\label{priorSect}

The value of the global smoothing parameter $\lambda$ determines the precision of the marginal distributions of the order-$k$ differences, which influences the smoothness of the estimated trend.  
Selection of the global smoothing parameter in penalized regression models is typically done via cross-validation in the frequentist setting \citep{tibshirani1996lasso} or marginal maximum likelihood in the empirical Bayes setting \citep{park2008bayesian}.  
Our fully Bayesian formulation eliminates the need for these additional steps, but in turn requires selection of the hyperparameter controlling the scale of the prior on the smoothing parameter.  
The value of this hyperparameter will depend on the order of the model, the grid resolution, and the variability in the latent trend parameters. 
Therefore, a single hyperparameter value cannot be used in all situations. 
Some recent studies have focused on methods for more careful and principled specification of priors for complex hierarchical models \citep{fong2010, simpson2014pc, sorbye2014}.      
The method of \cite{sorbye2014} was developed for intrinsic GMRF priors and we adapt their approach to our specific models in what follows.
\par

We wish to specify values of the hyperparameter $\zeta$ for various situations, where the global scale parameter $\gamma \sim \text{C}^+(0, \zeta)$ . 
Let $\boldsymbol{Q}$ be the precision matrix for the Markov random field corresponding to the model of interest (see Appendix \ref{appendixE} for examples), and $\boldsymbol{\Sigma} = \boldsymbol{Q}^{-1}$ be the covariance matrix with diagonal elements $\Sigma_{ii}$.  
The marginal standard deviation of all components of $\boldsymbol{\theta}$ for a fixed value of $\gamma$ is $ \sigma_\gamma(\theta_i) = \gamma \sigma_{\text{ref}}(\boldsymbol{\theta}) $, where  $\sigma_{\text{ref}}(\boldsymbol{\theta})$ is the geometric mean of the individual marginal standard deviations when $\gamma = 1$ \citep{sorbye2014}. 
We want to set an upper bound $U$ on the average marginal standard deviation of $\theta_i$, such that $ \Pr (\sigma_\gamma(\theta_i) > U ) = \alpha $, where $\alpha$ is some small probability.  
Using the cumulative probability function for a half-Cauchy distribution, we can find a value of $\zeta$ for a given value of $\sigma_{\text{ref}}(\boldsymbol{\theta})$ specific to a model of interest and given common values of $U$ and $\alpha$ by:   

\begin{equation}
  \zeta = \frac{U}{\sigma_{\text{ref}} (\boldsymbol{\theta})  \tan \left( \frac{\pi}{2}(1-\alpha)  \right)} .
   \label{zetafun} 
\end{equation}  

By standardizing calculations to be relative to the average marginal standard deviation, the methods of \cite{sorbye2014} allow us to easily calculate $\zeta$ for a model of different order or a model with a different density of grid points. 
For practical purposes we apply the same method to the normal and SPMRF models.  
This is not ideal in terms of theory, however, since the horseshoe distribution has infinite variance and the corresponding SPMRF will clearly not have the same marginal variance as a GMRF. 
This is not necessarily problematic since GMRF approximation will result in an estimate of $\zeta$ under the horseshoe SPMRF which is less informative than would result under similar methods derived specifically for the horseshoe SPMRF, and could therefore be seen as more conservative in terms of guarding against over smoothing.  
In contrast, the Laplace SPMRF has finite marginal variance that is well approximated by the GMRF methods.   
We apply these methods in the data examples that follow.

\subsection{Coal Mining Disasters}

This is an example of estimating the time-varying intensity of an inhomogeneous Poisson process that exhibits a relatively rapid period of change. The data are on the time intervals between successive coal-mining disasters, and were originally presented by \cite{maguire1952}, with later corrections given by \cite{jarrett1979} and \cite{raftery1986coal}.  We use the data format presented by \cite{raftery1986coal}.  A disaster is defined as an accident involving 10 or more deaths.  The first disaster was recorded in March of 1851 and the last in March of 1962, with 191 total event times during the period 1 January, 1851 through 31 December, 1962.  Visual inspection of the data suggests a decrease in rate of disasters over time, but it is unclear by eye alone whether this change is abrupt or gradual.  The decrease in disasters is associated with a few changes in the coal industry at the time. A sharp decline in labor productivity at the end of the 1880's is thought to have decreased the opportunity for disasters, and the formation of the Miner's Federation, a labor union, in late 1889 brought added safety and protection to the workers \citep{raftery1986coal}.
\par
This data set has been of interest to various authors due to uncertainty in the timing and rate of decline in disasters and the computational challenge presented by the discrete nature of the observations.  Some authors have fit smooth curves exhibiting gradual change \citep{adams2009tractable,teh2011gaussian} and others have fit change-point models with abrupt, instantaneous change \citep{raftery1986coal, carlin1992hierarchical, green1995}.  An ideal model would provide the flexibility to automatically adapt to either scenario.  
\par
We assumed an inhomogeneous Poisson process for the disaster events and binned the event counts by year.  
We fit first-order models using the normal, Laplace, and horseshoe prior formulations.  We assumed the event counts, $y_i$, were distributed Poisson conditional on the $\theta_i$: 
$ y_i \thinvert \theta_i \sim \text{Pois}\left(\exp(\theta_i)\right)$. 
The marginal prior distributions for the first-order increments were $\Delta\theta_j \sim \text{N}(0, \gamma^2)$ for the Normal,  $\Delta\theta_j \sim \text{Laplace}(\gamma)$ for the Laplace, and $\Delta\theta_j \sim \text{HS}(\gamma)$ for the horseshoe.  
We used the same prior specifications as those used in the simulations for the remaining parameters, except we used the guidelines in Section \ref{priorSect} to set the hyperparameter on the global scale prior. 
Using calculations outlined in Appendix \ref{appendixE}, we set $\sigma_{\text{ref}}(\boldsymbol{\theta})=6.47$ and $U=0.860$.  
Setting $\alpha=0.05$ and substituting into Equation (\ref{zetafun}) results in $\zeta = 0.0105$, so $\gamma \sim \text{C}^+(0, 0.0105)$ for each model.  
We used HMC for approximating the posterior distributions.  For each model we ran four independent chains, each with a burn-in of 500 followed by 6,250 iterations thinned at every 5.  This resulted in a total of 5,000 posterior samples for each model.  We were interested in finding the best representation of the process over time as well as finding the most likely set of years associated with the apparent change point. For this exercise we arbitrarily defined a change point as the maximum drop in rate between two consecutive time points.

\begin{figure}[t]
         \begin{center}
           \includegraphics[width=\textwidth]{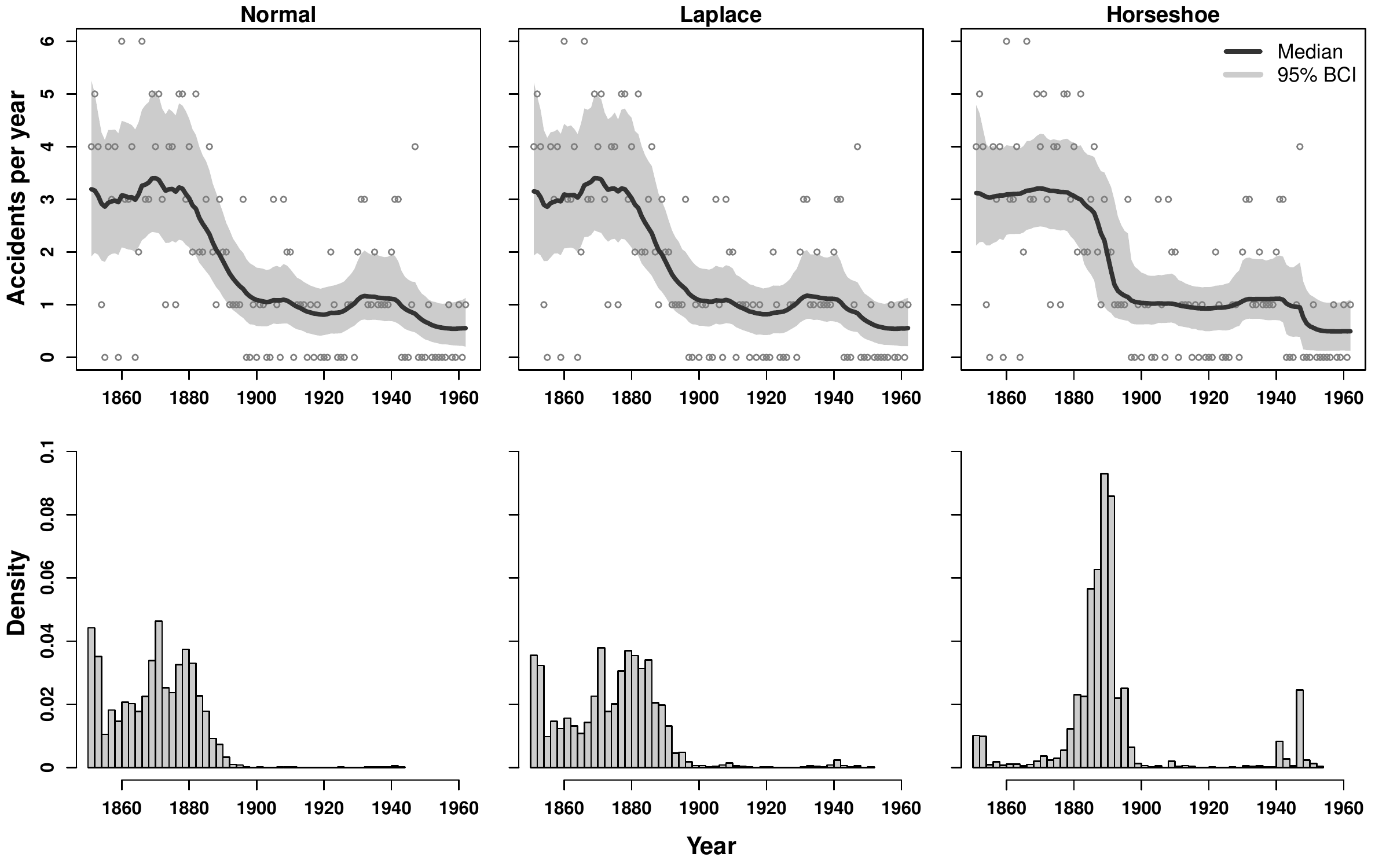}
         \end{center}
    \caption{Top row: fits to coal mining disaster data for different prior distributions. Posterior medians (lines), 95\% credible intervals (shaded regions), and data points are shown.  Bottom row: associated posterior distributions for change points. }
    \label{coalFig}
\end{figure} 
\par
Plots of the fitted trends (Figure \ref{coalFig}) indicate that the horseshoe model picked up a sharper change in trend and had narrower BCIs than the other models. 
The normal and Laplace models did not have sufficient flexibility to allow large jumps and produced a gradual decline in accidents rate, which 
is less plausible than a sharp decline in light of the additional information about change in coal mining industry safety regulations. 
The relative qualitative performance of the normal, Laplace, and horseshoe densities is  similar to that for the piecewise constant scenario from our simulation study.  
The posterior distributions of the change point times are shown in Figure \ref{coalFig}.  
The horseshoe model clearly shows a more concentrated posterior for the break points, and that distribution is centered near the late 1880's, which corresponds to the period of change in the coal industry.  
Therefore, we think the Bayesian trend filter with the horseshoe prior is a better default model in cases where sharp change points are expected.
\par
It is important to point out that we tried other values for the scale parameter ($\zeta$) in the prior distribution for $\gamma$ and found that the models were somewhat sensitive to that hyperparameter for this data set.  In particular, the horseshoe results for $\zeta = 1$ looked more like those for the other two models in Figure \ref{coalFig}, but when $\zeta = 0.0001$, the horseshoe produced more defined break points and straighter lines with narrower BCIs compared to the results with $\zeta = 0.01$ (see Appendix \ref{appendixF}).

\subsection{Tokyo Rainfall}

This problem concerns the estimation of the time-varying mean of an inhomogeneous binomial process.  We are interested in estimating the seasonal trend in daily probability of rainfall.  The data are binary indicators of when daily rainfall exceeded 1 mm in Tokyo, Japan, over the course of 39 consecutive years (1951-1989).  The indicators were combined by day of year across years, resulting in a sample size of $m=39$  for each of 365 out of 366 possible days, and a size of $m=10$ for the additional day that occurred in each of the 10 leap years.  The observation variable $y$ is therefore a count, where $y \in \{0, 1,\dots, 39\}$.  Data were obtained from the NOAA's National Center for Climate Information (https://www.ncdc.noaa.gov). A smaller subset of these data (1983-1984) was initially analyzed by \cite{kitagawa1987non} and later by several others, including \cite{rue2005gaussian}.

\begin{figure}[t] 
         \begin{center}
           \includegraphics[width=\textwidth]{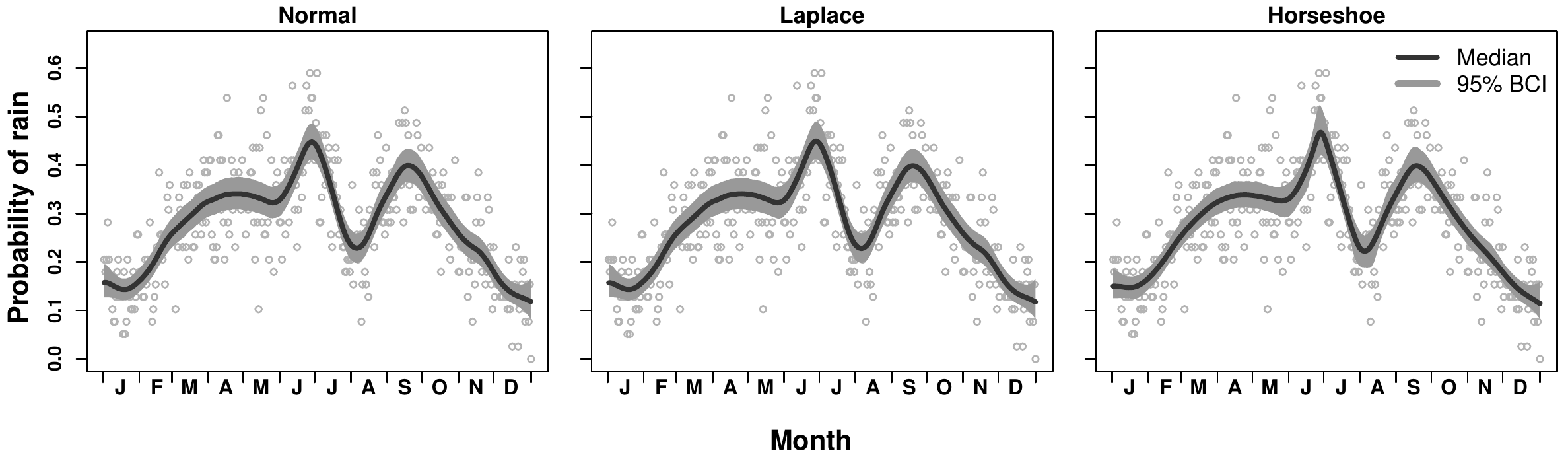}
         \end{center}
    \caption{Fits to Tokyo rainfall data for different prior distributions.  Posterior medians (lines), 95\% credible intervals (shaded regions), and estimated probabilities ($y_i/n_i$) are shown.  }
    \label{tokyofig}
\end{figure} 

We fit SPMRF models with Laplace and horseshoe priors and a GMRF model (normal prior).  
All models were based on second-order differences.  
The observation model was
\begin{equation*}
y_i \fatvert \theta_i \sim \text{Bin}\left(m_i, \,  \frac{1}{1 + \exp(-\theta_i) } \right),\\
\end{equation*} and the marginal prior distributions for the second-order differences were $\Delta^2\theta_j \sim \text{N}(0, \gamma^2)$ for the normal prior,  $\Delta^2\theta_j \sim \text{Laplace}(\gamma)$ for the Laplace, and $\Delta^2\theta_j \sim \text{HS}(\gamma)$ for the horseshoe.  
We used the same prior specifications as those used in the simulations for the remaining parameters, except we used the guidelines in Section \ref{priorSect} to set the hyperparameter on the global scale prior. 
Using calculations outlined in Appendix \ref{appendixE}, we set $\sigma_{\text{ref}}(\boldsymbol{\theta})=906.7$ and $U=0.679$.  
Setting $\alpha=0.05$ and substituting into Equation (\ref{zetafun}) results in $\zeta = 5.89 \times 10^{-5}$, so $\gamma \sim \text{C}^+(0, 5.89 \times 10^{-5})$ for each model.  
We ran four independent chains for each model, each with a burn-in of 500 followed by 6,250 draws thinned at every 5.  
This resulted in a total of 5,000 MCMC samples retained for each model. 
\par
The resulting function estimates for all models reveal a sharp increase in probability of rain in June followed by a sharp decrease through July and early August and a subsequent sharp increase in late August and September (Figure \ref{tokyofig}). 
Changes through the rest of the months were relatively smooth.  
The estimated function displays some variations in smoothness similar to the function with varying smoothness used in our simulations.  
All methods resulted in a similar estimated function, but the horseshoe prior resulted in a smoother function that displayed sharper features at transition points in late June and early August, yet also had narrower credible intervals over most of the function.  
The normal and Laplace models resulted in a little more variability in the trend in January-April and in November.  
In their analysis of a subset of these data, \cite{rue2005gaussian} used a circular constraint to tie together the endpoints of the function at the beginning and end of the year.  
We did not use such a constraint here, although it is possible with the SPMRF models. 
Even so, it is evident that the horseshoe model resulted in more similar function estimates at the endpoints than did the other two models.

\section{Discussion}

We presented a method for curve fitting in a Bayesian context that achieves locally adaptive smoothing by exploiting the sparsity-inducing properties of shrinkage priors and the smoothing properties of GMRFs.  
We compared the performance of the Laplace prior, which simply reformulates the frequentist trend filter to a Bayesian analog, to a more aggressive horseshoe shrinkage prior by using simulations and found that the horseshoe provided the best balance between bias and precision.  
The horseshoe prior has the greatest concentration of density near zero and the heaviest tails among the priors we investigated.  
This combination allows smooth functions to be fit in regions with weak signals or noisy data while still allowing for recovery of sharp functional changes when supported by informative data.  
The Laplace prior allowed more functional changes of moderate value to be retained and could not accommodate large changes without compromising the ability to shrink the noisy and smaller functional changes.  
This resulted in greater variability in the estimated functions and wider associated credible intervals for the models with the Laplace prior in comparison to those with the horseshoe prior when the underlying true functions had jumps or varying smoothness.  
The Laplace prior did have adaptive ability not possessed by the normal prior, but the horseshoe prior clearly had the best adaptive properties among the priors we investigated.
\par
The Laplace prior performed better than the horseshoe for the constant and smooth functions in our simulations, with results closer to those of the normal prior, although the differences in performance among the three methods were relatively small.  
These functions do not have large deviations in order-$k$ differences, and so there are many small or medium sized values for the estimated  $\Delta^k\theta$. 
This situation is reflective of cases described by \cite{tibshirani1996lasso} where the lasso and ridge regression perform best, which helps explain why the analogous SPMRF models with Laplace or normal prior distributions do better here.  
We expect that non-adaptive or mildly adaptive methods will perform better when used on functions which do not exhibit jumps or varying smoothness.  
However, it is reassuring that an adaptive method does nearly as well as a non-adaptive method for these functions.  
This allows an adaptive model such as that using the horseshoe to be applied to a variety of functions with minimal risk of performance loss.    
\par
Our fully Bayesian implementation of the SPMRF models eliminates the need to explicitly select the global smoothing parameter $\lambda$ , either directly or through selection methods such as cross-validation (e.g., \cite{tibshirani1996lasso}) or marginal maximum likelihood (e.g., \cite{park2008bayesian}). 
However, the fully Bayesian approach does still require attention to the selection of the hyperparameter that controls the prior distribution on the smoothing parameter. 
We found the methods of \cite{sorbye2014} to offer practical guidelines for selecting this hyperparameter, and we successfully applied a modification of those methods in our data examples.  
A highly informative prior on the global smoothing parameter can result in over-smoothing if the prior overwhelms the information in the data, while a diffuse prior may result in a rougher function with insufficient smoothing.  
Noisier data are therefore more sensitive to choice of parameterization of the prior on the global smoothing parameter. 
We tested prior sensitivity in the coal mining example and found that the horseshoe prior was more responsive to changes in hyperparmeter values than the normal and Laplace priors (see Appendix \ref{appendixF}).
However, the results for the simulations and for the Tokyo rainfall example were much more robust to the value of the hyperparameter on the global scale due to the information in the data.
As a precaution, we recommend first applying methods such as those by \cite{sorbye2014} to set the hyperparameter, but then also paying attention to prior sensitivity when analyzing noisy data with the SPMRF models. 
\par
We only addressed one-dimensional problems here, but we think the GMRF representation of these models can allow extension to higher dimensions such as the spatial setting by incorporating methods used by \cite{rue2005gaussian} and others.
We also plan to extend these methods to semi-parametric models that allow additional covariates.

\section{Acknowledgments}
%\begin{acknowledgement}
J.R.F, and V.N.M. were supported by the NIH grant R01 AI107034. J.R.F. was supported by the NOAA Advanced Studies Program and V.N.M. was supported by the NIH grant U54 GM111274.
We are grateful to Soumik  Pal for making us think harder about subordinated processes. We thank two anonymous reviewers for their help in improving this manuscript.
%\end{acknowledgement}

%\bibliographystyle{ba}
%\bibliographystyle{../biom} 
\bibliographystyle{ba} 
\bibliography{trendfilterBib}

% % % % % % % % % % % % % % % % % % % % % % %
% % %   APPENDIX
% % % % % % % % % % % % % % % % % % % % % %

\appendix
\counterwithin{figure}{section}
\counterwithin{table}{section}
\counterwithin{equation}{section}

\section{Approximation to the Horseshoe Density}
\label{appendixA}

There is no exact closed-form expression available for the horseshoe density function. We present an approximation to the horseshoe density that can be used without the need for explicit specification of the nuisance local scale parameters.  Following \cite{carvalho2010horseshoe}, the marginal distribution of $u$ given global scale parameter $\gamma$ is found by integrating over possible values of the local scale parameter $\tau$, where $u\thinvert\tau \sim \text{N}(0, \delta\tau^2) $ and $\tau \thinvert \gamma \sim \text{C}^+(0, \gamma)$.  Here $\delta$ is a constant representing a scale factor for the distance between adjacent points when this distribution is used for the increments of a $k$th-order smoothing model. This leads to

\begin{align*}
p(u  \thinvert \delta, \gamma) &= \int_{0}^{\infty} p(u  \thinvert \delta, \tau,  \gamma)p(\tau \thinvert \gamma) d\tau \nonumber \\
&=  \int_{0}^{\infty} \frac{1}{\sqrt{2\pi \delta \tau^2}} \exp\left(-\frac{u^2}{2\delta\tau^2}  \right) \frac{2\gamma}{\pi(\tau^2 + \gamma^2)}d\tau  
\end{align*}

\noindent We let $B = 2\gamma/(\sqrt{2\pi^3\delta}) $ and introduce the substitution $\omega = \tau^{-2}$, which gives $d\tau = -1/(2\omega^{3/2}) $, resulting in 

\begin{align*}
p(u  \thinvert \delta, \gamma) &= B  \int_{0}^{\infty} \frac{\omega^{1/2}} {2\omega^{3/2} } \exp\left(-\frac{ u^2\omega}{2\delta}  \right) \frac{1}{\omega^{-1} + \gamma^2} d\omega  \\  
&= \frac{B}{2}  \int_{0}^{\infty} \frac{1}{\omega} \exp\left(-\frac{u^2\omega}{2\delta}  \right) \frac{\omega}{1 + \omega\gamma^2} d\omega \\
&= \frac{B}{2}  \int_{0}^{\infty} \frac{1}{1 + \omega\gamma^2} \exp\left(-\frac{u^2\omega}{2\delta}  \right)  d\omega .
\end{align*}

\noindent Now we introduce the substitution $z = 1 + \omega\gamma^2  $, which gives $d\omega = \gamma^{-2}dz$, and results in  

\begin{align*}
p(u  \thinvert \delta, \gamma) &= \frac{B}{2}  \int_{1}^{\infty} \frac{1} {z\gamma^2} \exp\left\{ -\left(\frac{ u^2\omega}{2\delta}\right) \frac{z}{\gamma^2}  + \left( \frac{u^2}{2\delta}\right) \frac{1}{\gamma^2}  \right\}  dz  \\  
&= \frac{B}{2\gamma^2} \exp\left(\frac{u^2}{2\delta\gamma^2}  \right)  \int_{1}^{\infty} \frac{1} {z} \exp \left(-\frac{u^2 z}{2\delta\gamma^2}   \right) dz  \\
&= \left( \frac{1}{2\pi^3\delta\gamma^2} \right)^{1/2} \exp\left(\frac{u^2}{2\delta\gamma^2}  \right) \text{E}_1\left(\frac{u^2}{2\delta\gamma^2}  \right) dz  , 
\end{align*}

\noindent where $\text{E}_1$ is the exponential integral function.  Note that $\lim_{x\to 0^+} \text{E}_1(x) = \infty$, but for $x > 0$, the function $\text{E}_1(x)$ is bounded as follows:

\begin{equation*}
\frac{1}{2}e^{-x}\ln\left(1 + \frac{2}{x}\right) \, < \, \text{E}_1(x) \, < \, e^{-x}\ln\left(1 + \frac{1}{x}\right).
\end{equation*}  

\noindent  Then for $u \in \{\mathbb{R} : u \neq 0\}$ we have

\begin{equation*}
\frac{1}{2}\exp\left(\frac{-u^2}{2\delta\gamma^2}\right) \ln\left(1 + \frac{4\delta\gamma^2}{u^2}\right) \, < \, \text{E}_1\left(\frac{u^2}{2\delta\gamma^2}\right) \, < \, \exp\left(\frac{-u^2}{2\delta\gamma^2}\right) \ln\left(1 + \frac{2\delta\gamma^2}{u^2}\right).
\end{equation*}  

\noindent It follows that the target density is bounded by
\begin{equation}
\frac{1}{2}\left(\frac{1}{2\pi^3\delta\gamma^2}\right)^{1/2}\ln\left(1 + \frac{4\delta\gamma^2}{u^2}\right) \, < \, p(u  \thinvert \gamma) \, < \, \left(\frac{1}{2\pi^3\delta\gamma^2}\right)^{1/2}\ln\left(1 + \frac{2\delta\gamma^2}{u^2}\right). \label{bndhorse}
\end{equation}  
\noindent Let the left bound in equation (\ref{bndhorse}) be denoted $B_1(u)$ and the right bound $B_2(u)$. Note that as $u \rightarrow 0$, each of $B_1(u), p(u \thinvert \gamma)$ and $B_2(u)$ approach $\infty$. It can be shown that $\int_{-\infty}^{\infty}B_1(u)du = \sqrt{2/\pi}$ and $ \int_{-\infty}^{\infty}B_2(u)du = 2/\sqrt{\pi}$.  Since $\sqrt{2/\pi} < 1 < 2/\sqrt{\pi}$, these bounds can be used to find an approximate expression for $p(u \thinvert \gamma)$ that integrates to 1 and still satisfies equation (\ref{bndhorse}). We set \begin{equation}\tilde{p}(u \thinvert \gamma) =  w B_1(u) + (1-w)B_2(u) \label{appx1}\end{equation} with constraints $0 < w < 1$ and $\int_{-\infty}^{\infty} w B_1(u) + (1-w)B_2(u) du = 1 $.  Using the values for the integrated bounds and solving gives $w = (\sqrt{\pi}-2)/(\sqrt{2}-2) $. Substituting this value for $w$ into equation (\ref{appx1}) and simplifying gives the following closed-form approximation to the horseshoe density function:

\begin{equation}
\tilde{p}(u  \thinvert \gamma) = \left(\frac{1}{2\pi^3\delta\gamma^2}\right)^{1/2}\left[\frac{\sqrt{\pi}-2}{2\sqrt{2}-4} \ln\left(1 + \frac{4\delta\gamma^2}{u^2}\right)   + \frac{\sqrt{2}-\sqrt{\pi}}{\sqrt{2}-2}\ln\left(1 + \frac{2\delta\gamma^2}{u^2}\right) \right]. \label{apxhorse} 
\end{equation}

\clearpage

\section{Marginal Laplace Distribution with Irregular Grid Spacing}
\label{appendixB}

The following is a derivation of the marginal prior distribution for the order-$k$ differences when grid spacing is unequal.  These derivations are based on the scale-mixture representation of the Laplace distribution.  These results are known to apply to the first-order and second-order models, but higher orders.

Let $u_j = \Delta^k\theta_j$ and let $\delta_j$ be a constant representing a scale factor for the distance between adjacent points when this distribution is used for the increments of a $k$th-order smoothing model.  
For convenience, subscripts on $u$ and $\delta$ are dropped from here forward.  
We assume $u \vert \tau,\delta \sim \text{N}(0, \delta\tau^2) $ and $\tau^2 \vert \lambda^2 \sim \text{Exp}(\lambda^2/2)$.   Here $\lambda = 1/\gamma $ is the global shrinkage parameter.
It follows that 
\begin{align*}
p(u \vert \delta, \lambda) &=  \int_{0}^{\infty} \frac{\lambda^2}{2}\exp\left(-\frac{\tau^2\lambda^2}{2}\right) \frac{1}{\sqrt{2\pi\delta\tau^2}} \exp\left(-\frac{u^2}{2\delta\tau^2}\right) d\tau^2 \\
&= A  \int_{0}^{\infty}\frac{1}{\tau} \exp\left(-\frac{\tau^2\lambda^2}{2} -\frac{u^2}{2\delta\tau^2} \right)  d\tau^2 ,
\end{align*} where $A =  \frac{\lambda^2}{2\sqrt{2\pi\delta\tau^2}} $.  Now we make the substitution $\omega = 1/\tau^2$, which gives $d\tau^2 = -\omega^{-2} d\omega $, and the marginal density for $u$ becomes 

\begin{align*}
p(u \vert \delta, \lambda) &=  A  \int_{0}^{\infty}\omega^{-3/2} \exp\left\{-\frac{\lambda^2}{2\omega} -\frac{u^2\omega}{2\delta} \right\}  d\omega \\
&=  A  \int_{0}^{\infty}\omega^{-3/2} \exp\left\{-\frac{u^2\omega}{2\delta} -\frac{\lambda^2}{2\omega} + \frac{\lambda\vert u \vert}{\delta^{1/2}} - \frac{\lambda\vert u \vert}{\delta^{1/2}} \right\}  d\omega \\
&=  A  \int_{0}^{\infty}\omega^{-3/2} \exp\left\{  -\frac{|u|^2}{2\delta\omega} \left( \omega^2 - \frac{2\delta^{1/2}\omega\lambda}{|u|} + \frac{\delta\lambda^2}{|u|^2} \right) - \frac{\lambda\vert u \vert}{\delta^{1/2}} \right\}  d\omega  \\
&=  \frac{\lambda}{2\sqrt{\delta}} \exp\left\{ - \frac{\lambda\vert u \vert}{\sqrt{\delta} }  \right\}  \int_{0}^{\infty} \frac{\lambda}{\sqrt{2\pi}\omega^{3/2}}  \exp\left\{   -\frac{\lambda^2}{2\delta\omega(\lambda^2/|u|^2)} \left( \omega -  \frac{\sqrt{\delta}\lambda}{|u|} \right)^2  \right\}  d\omega \\
&=  \frac{\lambda}{2\sqrt{\delta}} \exp\left\{ - \frac{\lambda\vert u \vert}{\sqrt{\delta} }  \right\} , 
\end{align*}
where the last line follows from the fact that the integrand in the second-to-last line is the pdf of an inverse-Gaussian distribution with mean parameter $\mu = \sqrt{\delta}\lambda/|u| $ and shape parameter $\lambda = \lambda^2$.  The result is the pdf of a Laplace distribution with mean zero and scale parameter $\lambda/\sqrt{\delta}$.  Note that the variance of the Laplace distribution is $2\delta/\lambda^2$, which implies that the grid spacing $\delta$ scales the variance of the increments $u$.

\section{Data Example with Irregular Grid}
\label{appendixC}

We apply the SPMRF models to a data set with a continuous covariate.  
The response data are rent per square meter of floor space in Munich, Germany, and the covariate is the floor space in square meters.  
These data were analysed by \cite{rue2005gaussian} using a second-order GMRF with irregular spacing.  
Here we apply a second-order GMRF and SPMRF models using the methods described in Section \ref{methodIrreg} of the main text.  
\par 
Let $x$ represent the floor space measurements, and let $x_1 < x_2 < ... < x_n$ be the ordered set of unique floor measurement values.
Further, let $\delta_j = x_{j+1} - x_{j}$ be the distance between adjacent floor space measurements.  
The marginal prior distributions for the second-order differences were $\Delta^2\theta_j \sim \text{N}(0, d_j\gamma^2)$ for the normal prior,  $\Delta^2\theta_j \sim \text{Laplace}(d_j^{1/2}\gamma)$ for the Laplace, and $\Delta^2\theta_j \sim \text{HS}(d_j^{1/2}\gamma)$ for the horseshoe, where 
$$ \Delta^2\theta_{j} = \theta(x_{j+2}) - \left( 1+ \frac{\delta_{j+1}}{\delta_j} \right) \theta(x_{j+1}) + \frac{\delta_{j+1}}{\delta_j}\theta(x_{j}), $$ 
and 
$$ d_j = \frac{\delta_{j+1}^2(\delta_j+\delta_{j+1})}{2}. $$
Using methods described in Section \ref{priorSect} of the main text and Appendix \ref{appendixE}, we calculated the value of the hyperparameter for the global scale parameter to be $\zeta = 0.00094$, so $\gamma \sim \text{C}^+(0, 0.00094) $ for all models.  The results are shown in Figure \ref{rentfig}.

\bigskip
\bigskip
\bigskip
\bigskip

\begin{figure}[h]
	\begin{center}
		\includegraphics[width=\textwidth]{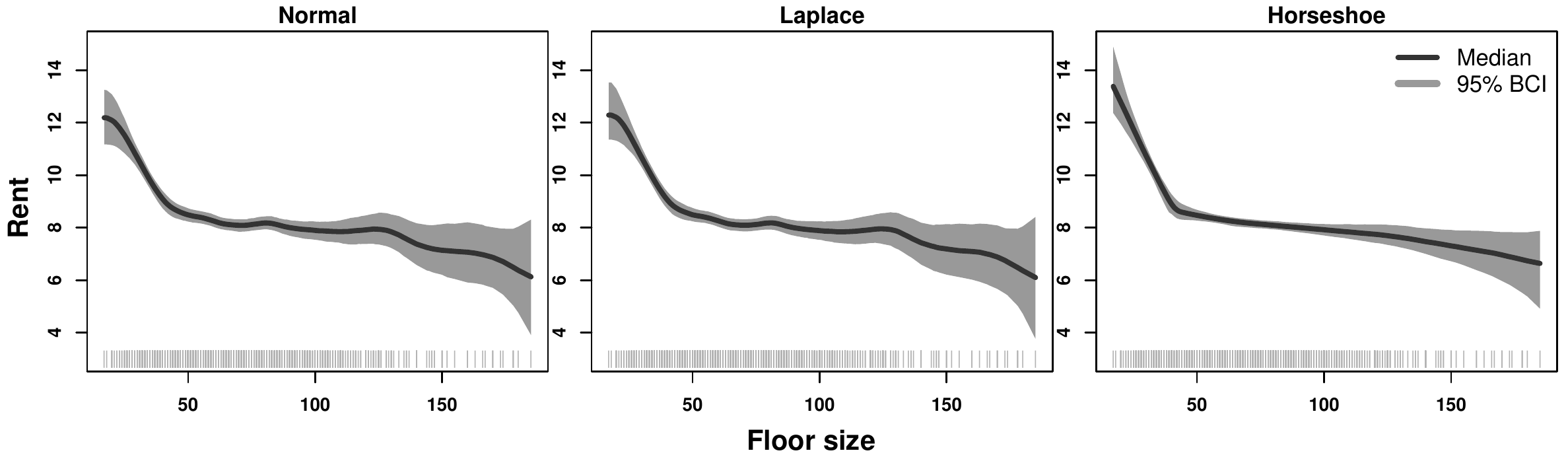}
	\end{center}
	\caption{Results for models using irregular grids for Munich rent data.  Posterior medians (dark line) are shown with 95\% Bayesian credible intervals (BCIs).  Locations of data are shown with vertical bars at the bottom of plots.  \label{rentfig}}
\end{figure} 

\clearpage

\section{Additional Simulation Results}
\label{appendixD}

Here we display plots with simulation results for normal data with $\sigma=1.5$ (Figure \ref{simresultsSN6}), Poisson data (Figure \ref{simresultsP}), and binomial data (Figure \ref{simresultsBin}).  Summary measures for all data types show similar patterns to each other and to those for normal data with $\sigma=4.5$ (Figure 2 in main article).

\begin{table}[h]
\caption{Mean values of performance measures across 100 simulations for normal observations ($\sigma=1.5$) for each model and trend function type. } \label{normtab6}
\centering
\begin{tabular}{llrrrr}
    %\hline
    \toprule
 \textbf{Function} & \textbf{Model} & \textbf{MAD} & \textbf{MCIW} & \textbf{MASV} & \textbf{TMASV}  \\ 
    %\hline
    \midrule
Constant & Normal& 0.115 & 0.698 & 0.002 & 0.000 \\ 
   & Laplace  & 0.116 & 0.731 & 0.002 & 0.000 \\ 
   & Horseshoe  & 0.127 & 0.921 & 0.004 & 0.000 \\ 
    %\hline
    \midrule
Piecewise Const. & Normal  & 1.040 & 5.479 & 1.647 & 0.606 \\ 
   & Laplace  & 0.899 & 3.282 & 1.557 & 0.606 \\ 
   & Horseshoe & 0.281 & 1.918 & 0.638 & 0.606 \\ 
    %\hline
    \midrule
Smooth & Normal  & 0.565 & 2.985 & 1.391 & 1.406 \\ 
   & Laplace  & 0.561 & 2.985 & 1.393 & 1.406 \\ 
   & Horseshoe & 0.565 & 2.946 & 1.414 & 1.406 \\ 
    %\hline
    \midrule
Varying Smooth & Normal & 0.586 & 3.036 & 0.613 & 0.543 \\ 
   & Laplace & 0.550 & 2.898 & 0.592 & 0.543 \\ 
   & Horseshoe  & 0.438 & 2.228 & 0.558 & 0.543 \\ 
    % \hline
    \bottomrule
\end{tabular}
\end{table}

  \begin{table}[h]
 \caption{Mean values of performance measures across 100 simulations for Poisson observations for each model and trend function type. } \label{poistab}
 \centering
 \begin{tabular}{llrrrr}
    %\hline
    \toprule
 \textbf{Function} & \textbf{Model} & \textbf{MAD} & \textbf{MCIW} & \textbf{MASV} & \textbf{TMASV}  \\ 
    %\hline
    \midrule
 Constant & Normal  & 0.022 & 0.142 & 0.001 & 0.000 \\ 
    & Laplace & 0.023 & 0.149 & 0.001 & 0.000 \\ 
    & Horseshoe  & 0.025 & 0.167 & 0.001 & 0.000 \\ 
    %\hline
    \midrule
 Piecewise Const. & Normal & 0.109 & 0.557 & 0.077 & 0.030 \\ 
    & Laplace & 0.092 & 0.529 & 0.064 & 0.030 \\ 
    & Horseshoe  & 0.051 & 0.334 & 0.036 & 0.030 \\ 
    %\hline
    \midrule
 Smooth & Normal   & 0.078 & 0.379 & 0.072 & 0.079 \\ 
    & Laplace  & 0.078 & 0.380 & 0.072 & 0.079 \\ 
    & Horseshoe & 0.079 & 0.382 & 0.073 & 0.079 \\ 
    %\hline
    \midrule
 Varying Smooth & Normal & 0.067 & 0.296 & 0.020 & 0.023 \\ 
    & Laplace  & 0.066 & 0.295 & 0.020 & 0.023 \\ 
    & Horseshoe & 0.058 & 0.277 & 0.020 & 0.023  \\ 
    % \hline
    \bottomrule
 \end{tabular}
 \end{table}

 \begin{table}[h]
  \caption{Mean values of performance measures across 100 simulations for binomial observations for each model and trend function type. } \label{bintab}
  \centering
  \begin{tabular}{llrrrr}
    %\hline
    \toprule
 \textbf{Function} & \textbf{Model} & \textbf{MAD} & \textbf{MCIW} & \textbf{MASV} & \textbf{TMASV}  \\ 
    %\hline
    \midrule
    %\endhead
  Constant & Normal  & 0.042 & 0.249 & 0.001 & 0.000 \\ 
     & Laplace & 0.043 & 0.262 & 0.001 & 0.000 \\ 
     & Horseshoe  & 0.047 & 0.311 & 0.002 & 0.000 \\ 
    %\hline
    \midrule
  Piecewise Const. & Normal & 0.229 & 1.191 & 0.166 & 0.066 \\ 
     & Laplace & 0.193 & 1.126 & 0.137 & 0.066 \\ 
     & Horseshoe  & 0.108 & 0.690 & 0.076 & 0.066\\ 
    %\hline
    \midrule
  Smooth & Normal   & 0.139 & 0.733 & 0.110 & 0.117\\ 
     & Laplace  & 0.139 & 0.735 & 0.111 & 0.117  \\ 
     & Horseshoe & 0.143 & 0.740 & 0.113 & 0.117 \\ 
     %\hline
     \midrule
  Varying Smooth & Normal & 0.188 & 0.730 & 0.056 & 0.068 \\ 
     & Laplace  & 0.183 & 0.726 & 0.056 & 0.068 \\ 
     & Horseshoe  & 0.149 & 0.676 & 0.058 & 0.068 \\ 
    % \hline
    \bottomrule
  \end{tabular}
 \end{table}

\begin{figure}[h]
	\begin{center}
		\includegraphics[width=\textwidth]{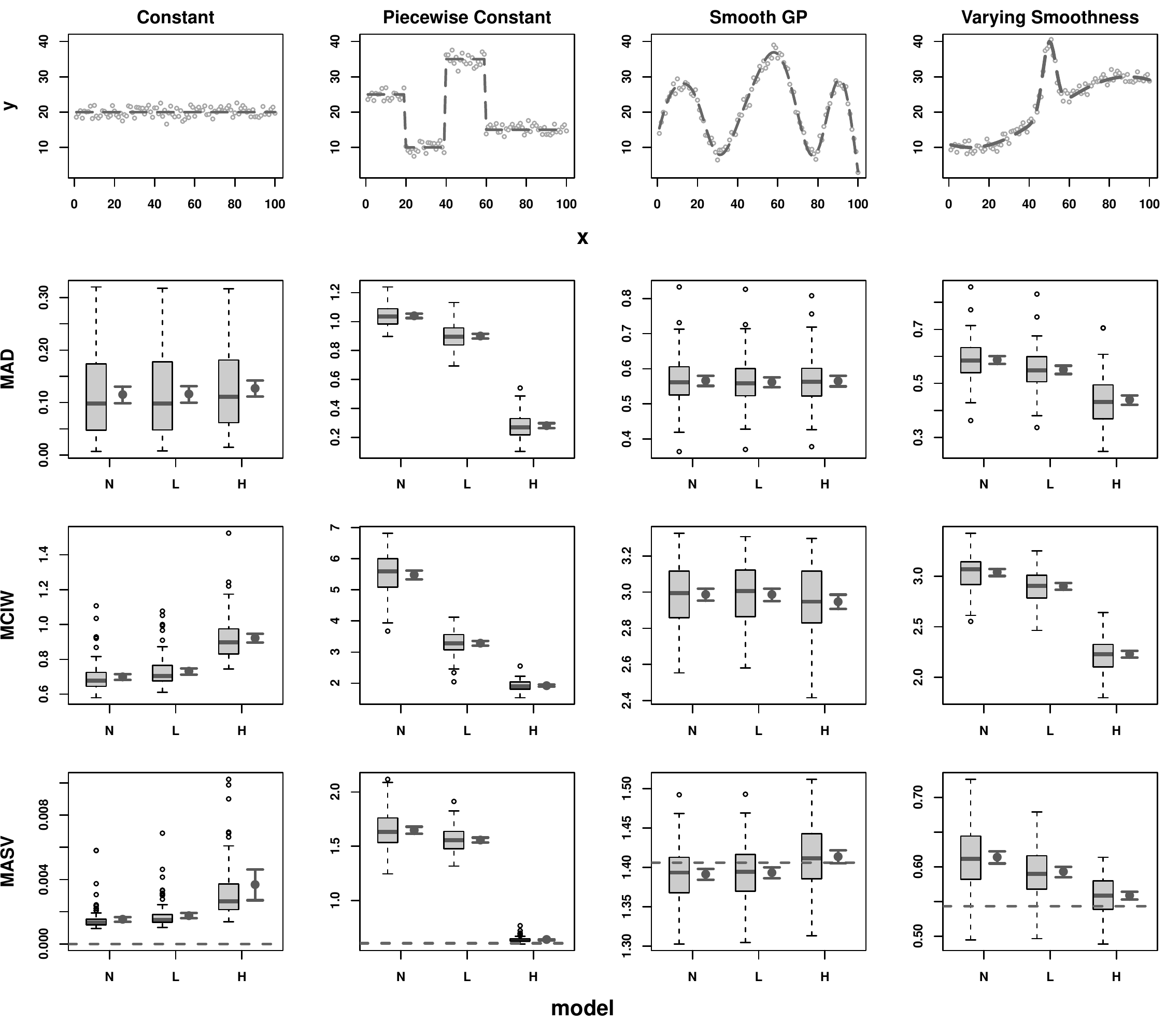}
	\end{center}
	\caption{Functions used in simulations and simulation results by model (N=normal, L=Laplace, H=horseshoe) and function type for normally distributed data with $\sigma$ = 1.5. Top row shows true functions (dashed lines) with example simulated data.  Remaining rows show mean absolute deviation (MAD), mean credible interval width (MCIW), and mean absolute sequential variation (MASV).  Horizontal dashed line in plots on bottom row is the true mean absolute sequential variation (TMASV). Shown for each model are standard boxplots of simulation results (left) and mean values with 95\% frequentist confidence intervals (right).  \label{simresultsSN6}  }
\end{figure}

\begin{figure}[h]
	\begin{center}
		\includegraphics[width=\textwidth]{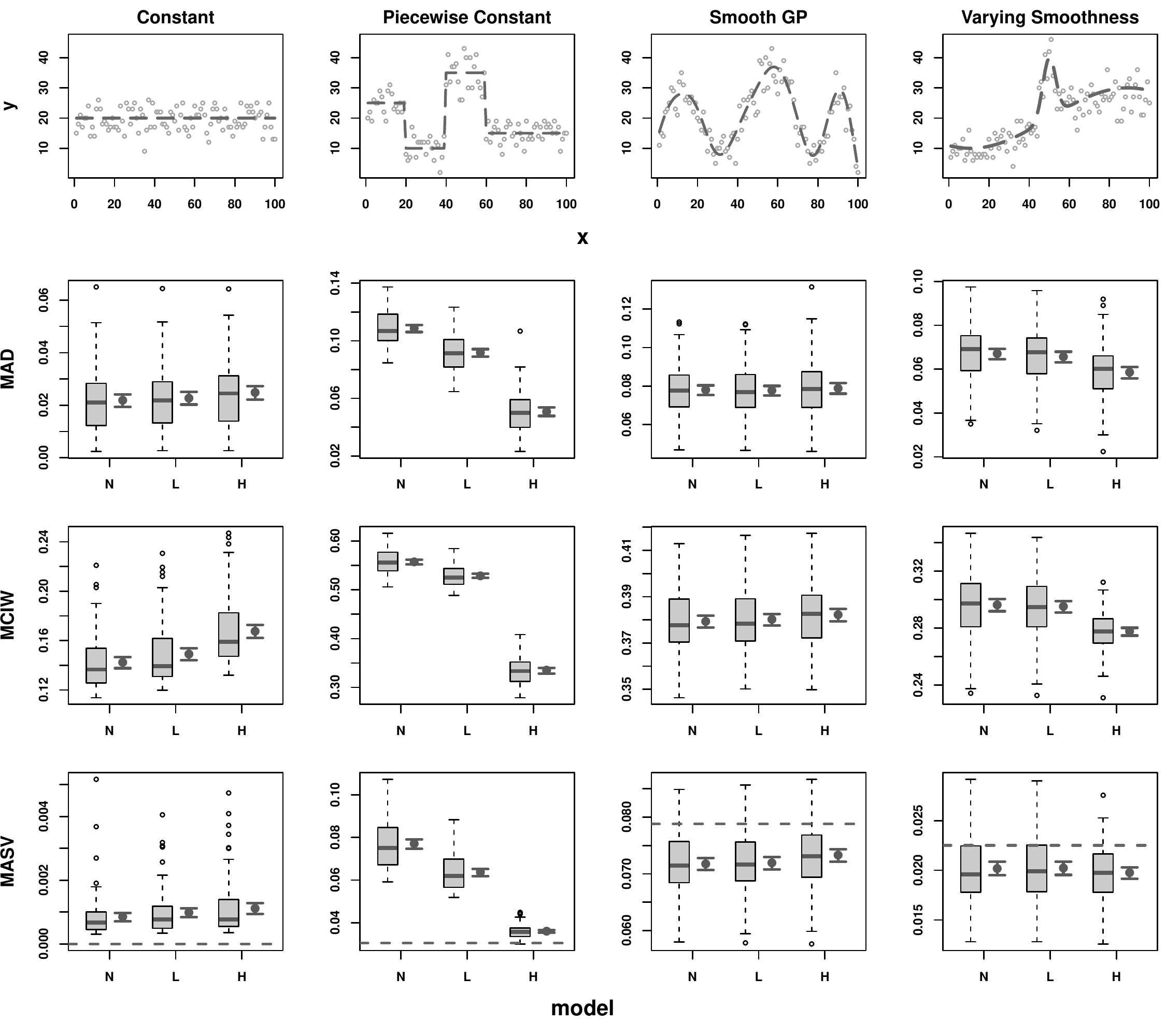}
	\end{center}
	\caption{Functions used in simulations and simulation results by model (N=normal, L=Laplace, H=horseshoe) and function type for Poisson distributed data. Top row shows true functions (dashed lines) with example simulated data.  Remaining rows show mean absolute deviation (MAD), mean credible interval width (MCIW), and mean absolute sequential variation (MASV).  Horizontal dashed line in plots on bottom row is the true mean absolute sequential variation (TMASV). Shown for each model are standard boxplots of simulation results (left) and mean values with 95 \% frequentist confidence intervals (right).  \label{simresultsP} }  
\end{figure}

\begin{figure}[h]
	\begin{center}
		\includegraphics[width=\textwidth]{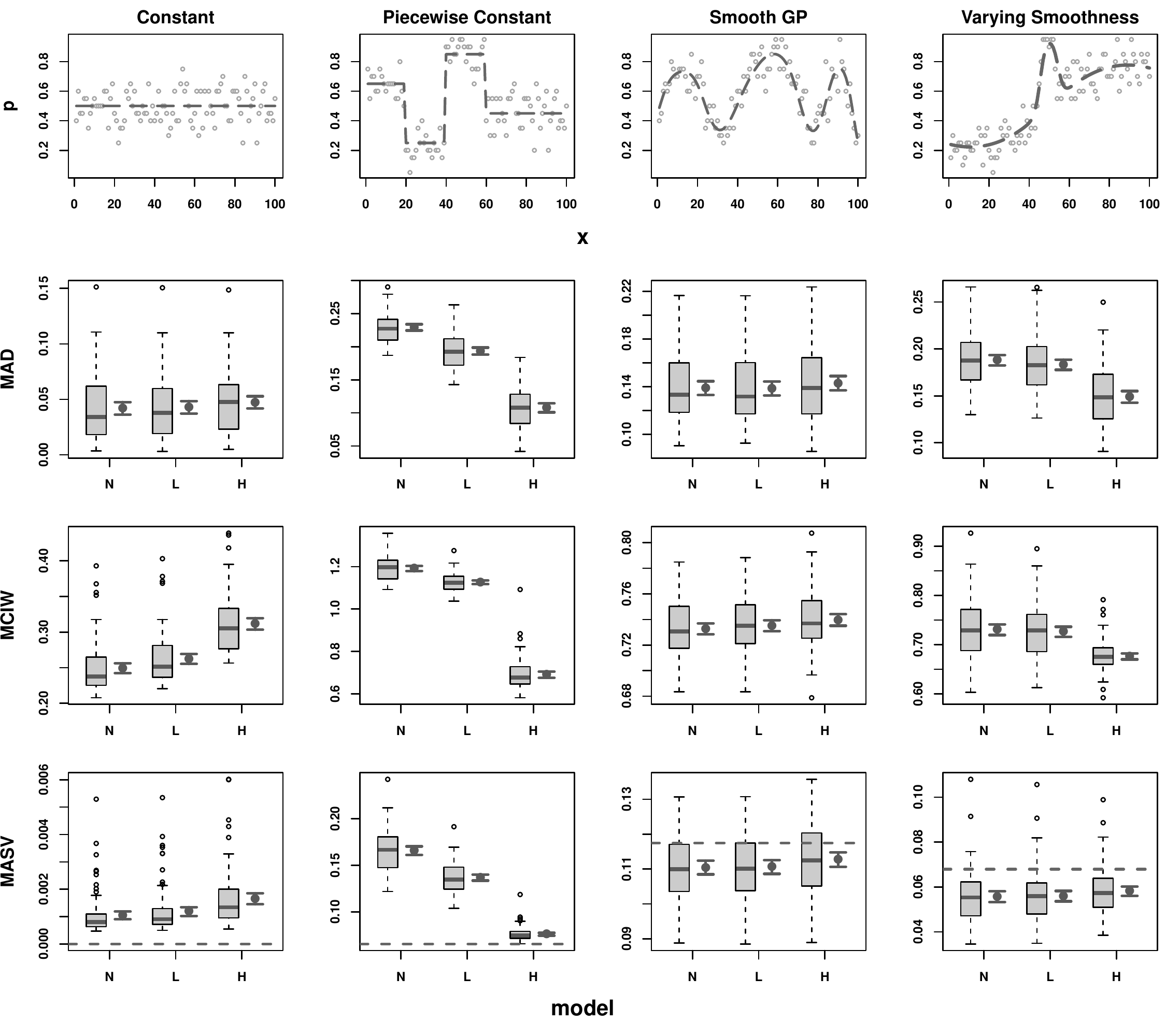}
	\end{center}
	\caption{Functions used in simulations and simulation results by model (N=normal, L=Laplace, H=horseshoe) and function type for binomial distributed data. Top row shows true functions (dashed lines) with empirical probability estimates from example simulated data.  Remaining rows show mean absolute deviation (MAD), mean credible interval width (MCIW), and mean absolute sequential variation (MASV).  Horizontal dashed line in plots on bottom row is the true mean absolute sequential variation (TMASV). Shown for each model are standard boxplots of simulation results (left) and mean values with 95\% frequentist confidence intervals (right).   \label{simresultsBin} }  
\end{figure}

\clearpage

\section{Parameterizing the Global Smoothing Prior}
\label{appendixE}

Here we provide additional details for calculating the hyperparameter $\zeta$ for the prior on the global scale parameter $\gamma$, where $\gamma \sim \text{C}^+(0, \zeta)$. 
First let $\boldsymbol{Q}$ be the precision matrix for the Markov random field corresponding to the model of interest (see examples below), and $\boldsymbol{\Sigma} = \boldsymbol{Q}^{-1}$ be the covariance matrix with diagonal elements $\Sigma_{ii}$.  
Following \cite{sorbye2014}, the marginal standard deviation of all components of $\boldsymbol{\theta}$ for a fixed value of $\gamma$ is $ \sigma_\gamma(\theta_i) = \gamma \sigma_{\text{ref}}(\boldsymbol{\theta}) $, where
\begin{equation}
	\sigma_{\text{ref}}(\boldsymbol{\theta}) = \exp \left( \frac{1}{n}\sum_{i=1}^{n}\log \sigma_{\{\gamma=1\}}(\theta_i)  \right) = \exp \left(\frac{1}{n}\sum_{i=1}^{n}\frac{1}{2}\log \left(\Sigma_{ii} \right)  \right) .
\end{equation}
That is, $\sigma_{\text{ref}}(\boldsymbol{\theta})$ is the geometric mean of the individual marginal standard deviations when $\gamma = 1$. 
\par

Before going further, let us describe two precision matrices used in the accompanying paper and their associated covariance matrices. \cite{sorbye2014} use intrinsic formulations of a Gaussian Markov random field (GRMF), which is also possible with our models, but we chose to use proper GMRF models.  This requires specification of the variance of the first $\theta$, which we denote $\omega^2 = \text{Var}(\theta_1)$.  For a given value of $\gamma$, the $n\times n$ precision matrix for a first-order model is:
\begin{equation}
  Q_1 = 1/\gamma^2 \begin{pmatrix}
  \gamma^2/\omega^2+1 & -1 &            & & & \\
  -1         &  2 & -1 & & & \\ 
                       & -1 &  2 & -1 & \\
                       &             &    \ddots   &  \ddots    & \ddots \\
                       &             &             &  -1 &  2 & -1  \\
                       &             &             &              & -1 & 1  
  \end{pmatrix},
\end{equation}
and the corresponding covariance matrix is:
\begin{equation}
\Sigma_1 = \begin{pmatrix}
\omega^2 & \omega^2            &  \cdots             &                     &         & \omega^2  \\
\omega^2 & \omega^2 + \gamma^2 & \omega^2 + \gamma^2 & \cdots               &         & \omega^2 + \gamma^2 \\ 
\vdots & \omega^2 + \gamma^2 & \omega^2 + 2\gamma^2 & \omega^2 + 2\gamma^2 & \cdots & \omega^2 + 2\gamma^2 \\
&  \vdots             & \omega^2 + 2\gamma^2   &  \ddots            &         & \vdots \\
&                     &     \vdots             &                   & \omega^2 + (n-2)\gamma^2        & \omega^2 + (n-2)\gamma^2  \\
\omega^2 &  \omega^2 + \gamma^2 & \omega^2 + 2\gamma^2       & \cdots    & \omega^2 + (n-2)\gamma^2 & \omega^2 + (n-1)\gamma^2  
\end{pmatrix}.
\end{equation}
\noindent Therefore, the marginal variances for the first-order model are $\Sigma_{1,ii} = \omega^2 + (i-1)\gamma^2$. 
For the second-order model, the $n \times n$ precision matrix is: 
\begin{equation}
Q_2 = 1/\gamma^2 \begin{pmatrix}
   \gamma^2/\omega^2+2 & -3 &  1 &         &   &  & & &\\
  -3                   &  6 & -4 & 1 &     &   &  & & &\\ 
    1                  & -4 &  6 & -4 &  1 &   &  & & &\\
                       & 1  & -4 &  6 & -4 & 1 &  & & &\\
   & & \ddots             &  \ddots     &    \ddots    &  \ddots    & \ddots & & \\
   & &   &                1  &       -4 &  6 & -4 & 1 &  \\
   & &   & &                1  &       -4 &  6 & -4 & 1  \\
   & &   & &                   &       1  &  -4 &  5 & -2  \\
   & & & &                   &             &  1 &  -2 & 1 
\end{pmatrix}.
\end{equation}
There is an analytical form for the covariance matrix for the second-order model, but it suffices here to know that the form of the marginal variances is:
\begin{equation}
  \Sigma_{2,ii} = \omega^2 + \frac{i(i-1)(2i-1)}{6}\gamma^2.
\end{equation} 
Note that if we were using an intrinsic GMRF model, we would assume that $\omega^2$ is infinite, which would result in a covariance matrix of rank $n-k$. Following \cite{sorbye2014} we would then use the generalized inverse of the precision matrix to calculate the marginal variances.

In practice we use the variance of the data (transformed data if the $\theta$ parameters are on a transformed scale) as an estimate of $\omega^2$.  Although this is using the data twice, this offers a reasonable constraint on the marginal variances of the $\theta$s. 

We want to set an upper bound $U$ on the average marginal standard deviation of $\theta_i$, such that $ \Pr (\sigma_\gamma(\theta_i) > U ) = \alpha $, where $\alpha$ is some small probability.  
Using the cumulative probability function for a half-Cauchy distribution, we can find a value of $\zeta$ for a given value of $\sigma_{\text{ref}}(\boldsymbol{\theta})$ specific to a model of interest and given common values of $U$ and $\alpha$ by:  
\begin{equation}
	\zeta = \frac{U}{\sigma_{\text{ref}} (\boldsymbol{\theta})  \tan \left( \frac{\pi}{2}(1-\alpha)  \right)} .
	\label{zetafun} 
\end{equation}  
It may be useful to note here that the median of a half-Cauchy distribution is equal to its scale parameter, since the median may be a more intuitive measure of the effect of different values of $\zeta$. 
\par
For our data examples in the main text, we let $U$ be the estimated standard deviation of the data on the appropriate scale.  We know that the marginal variances of the $\theta$s should not exceed the variance in the observed data, on average.  We set $\alpha=0.05$ as the probability of the average marginal standard deviation exceeding $U$.  For the coal mining example in the main text, we found an estimate of the variance of the data on the log scale by $\sum_{i=1}^{n}\ln(y_i+0.5)/(n-1)$, where $y_i$ is the observed count at time $i =1,\dots,n$.  For the Tokyo rain example, we estimated the variance of the data on the logit scale as  $\sum_{i=1}^{n}\text{logit}((y_i+q_i)/m_i))/(n-1) $, where $y_i$ is the number of years with rain on day $i$ out of $m_i$ possible years, and $q_i = 0.005I_{y_i=0} - 0.005I_{y_i=1} + 0I_{y_i \notin \{0,1\}}  $, where $I$ is an indicator function.  

Suppose we have calculated $\zeta_{o1}$, the hyperparameter for a first-order model given the corresponding average marginal standard deviation $\sigma_{\text{ref}}(\boldsymbol{\theta}_{o1})$ using Equation (\ref{zetafun}).  
If we wish to calculate the value of $\zeta_{o2}$ for a second-order model we can simply use
$$ \zeta_{o2} = \zeta_{o1} \frac{\sigma_{\text{ref}}(\boldsymbol{\theta}_{o1}) } {\sigma_{\text{ref}}(\boldsymbol{\theta}_{o2}) } .$$ 
Now suppose we have a model with $n$ equidistant nodes and want to increase the density of the grid to $kn$ nodes without changing the range of the grid.  
For a first-order model, $\text{Var}(\Delta\theta_{\text{new}}) = \frac{1}{k} \text{Var}(\Delta\theta) $, and for a second-order model $\text{Var}(\Delta^2\theta_{\text{new}}) = \frac{1}{k^3} \text{Var}(\Delta^2\theta)$ \citep{lindgren2008second, sorbye2014}.  
In terms of the hyperparameter for the global smoothing prior, for the first-order model $\zeta_{o1,\text{new}} = k^{-1/2}\zeta_{o1}$, and for the second-order model $\zeta_{o2,\text{new}} = k^{-3/2}\zeta_{o2}$.

\clearpage

\section{Prior Sensitivity}
 \label{appendixF}

We tested the sensitivity of the three prior formulations (normal, Laplace, and horseshoe) to the value of the hyperparameter ($\zeta$) which controls the scale of the distribution on the smoothing parameter $\gamma$, where $\gamma \sim \text{C}^+(0,\zeta)$  A smaller value of $\zeta$ constricts $\gamma$ to be closer to zero, which in turn constricts the scales of the priors on the order-$k$ differences.  We tested three levels for the hyperparameter: a) $\zeta=1$, b) $\zeta=0.01$, and c) $\zeta=0.0001$.  In general, we expect noisier data sets should be more sensitive to prior settings.  The coal mine disaster data offered a good test set because the observations are relatively noisy.   
\par
Clearly the horseshoe prior was the most sensitive to the level of $\zeta$ (Figure \ref{coalSens}). 
In particular, the horseshoe results for $\zeta = 1$ looked more like those for the other two models in Figure \ref{coalFig}, but when $\zeta = 0.0001$, the horseshoe produced more defined break points and straighter lines with narrower BCIs compared to the results with $\zeta = 0.01$.

\begin{figure}
	\begin{center}
		\includegraphics[width=\textwidth]{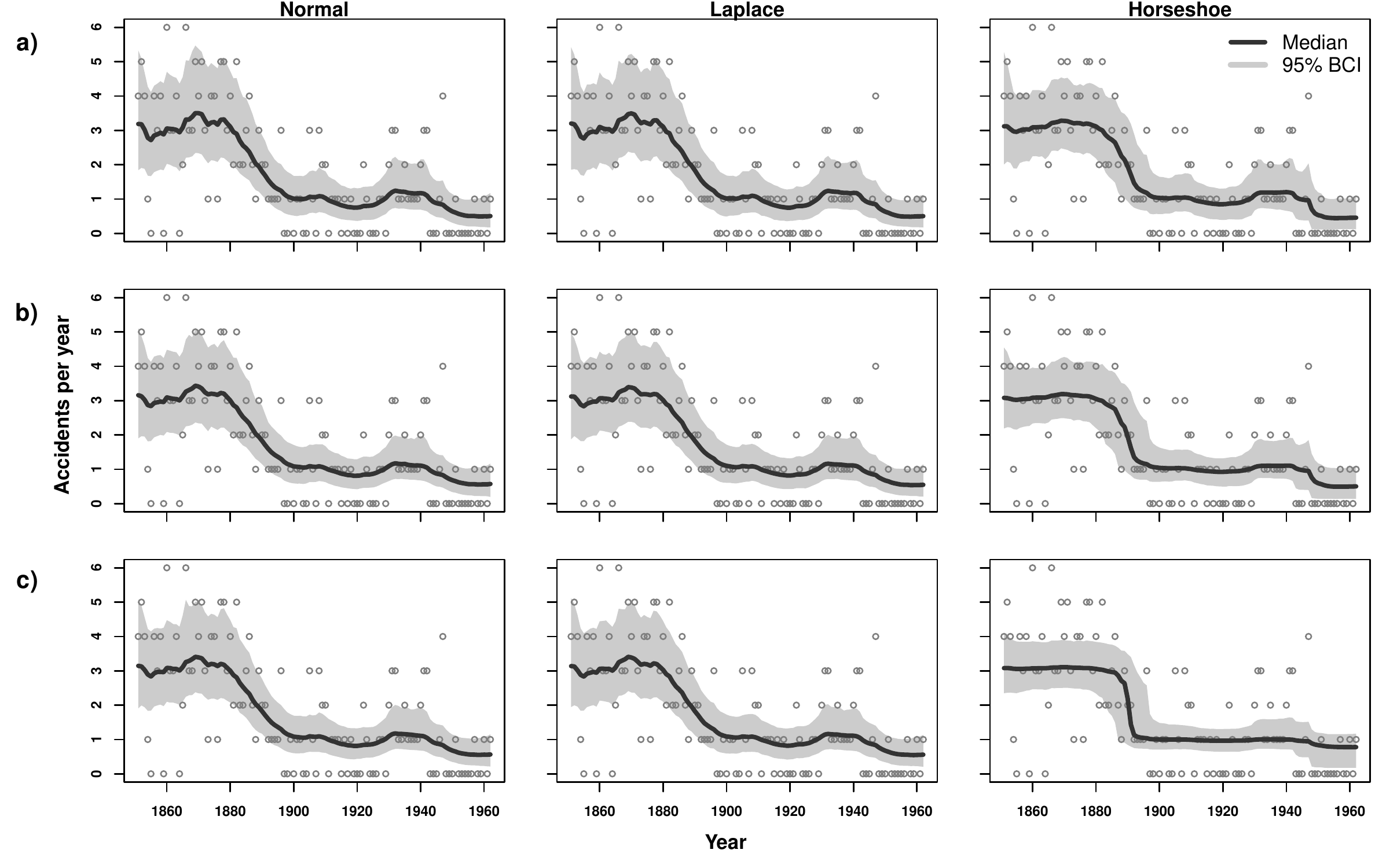}
	\end{center}
	\caption{Models fits to coal mining accidents data by model type and value of hyperparameter for global scale: a) $\zeta=1$, b) $\zeta=0.01$, and c) $\zeta=0.0001$. Posterior medians and associated 95\% Bayesian credible intervals are shown along with observed data.    \label{coalSens}  }
\end{figure}

\section{Computational Efficiency}
 \label{appendixG}

To compare SPMRF and GMRF models' computational efficiency, we calculated the effective number of posterior samples per second of computation time (ESSps)  for different model formulations and data configurations.  
We used the scenario with a piecewise-constant expected value from our main simulations (Section 3) to test the effect of model type, model order, and number of grid cells ($n$) on the ESS per second of sampling time.  
Here sampling time is defined as the total run time minus the time spent in the adaptation (warm-up or burn-in) phase, where time is measured in seconds of CPU time.  
We also calculated the ESSps for the coal mining and Tokyo rainfall examples.
\par
There were three simulated scenarios with piecewise-constant trend: 1) order-1 with $n=100$ observations and grid points (one observation per grid point), 2) order-1 with $n=200$, and 3) order-2 with $n=100$.  
The observations in these scenarios were normally distributed with standard deviation $\sigma=4.5$.
For each of these scenarios we ran 4 independent chains each with 1,000 iterations of burn-in and 2,500 iterations post burn-in thinned at every 5 for a total of 2,000 retained posterior samples combined across chains.  
The maximum ESS would therefore be 2,000 for these scenarios.  
The chains were run in sequence so that the total time (TCPU) sampling times (SCPU) times are the respective cumulative times across 4 chains.  
For the coal mining and Tokyo rainfall examples we used the same settings for number of iterations and thinning as was used in the main text (Section 4).  
We calculated effective sample size using the methods described in the documentation for \texttt{stan} \citep{stan-manual:2015}. 
\par
  \begin{table}[h]
 \caption{Measures of computational efficiency for each model type (Normal (N), Laplace (L), or Horseshoe (H)) for three simulated data scenarios and two real data examples.  Model order and number of parameters ($p$) are shown. The total CPU time (TCPU: includes adaptive phase) and sampling CPU (SCPU) time are in seconds.  The minimum and mean effective sample sizes per second (ESSps) of SCPU are also shown.} \label{esstab}
 \centering

\begin{longtable}{LCCCRRSS}
   \toprule
\textbf{Scenario} & \textbf{Model} & \textbf{Order} & \textbf{\textit{p}} & \textbf{TCPU} & \textbf{SCPU} & \textbf{Min. ESSps} & \textbf{Mean ESSps}  \\
\midrule
\endhead
Piece. Const. ($n=100$) & N & 1 & 102 & 74 & 52 & 25.14 & 36.79  \\
                           & L & 1 & 201 & 422 & 290 & 5.13 & 6.58  \\
                           & H & 1 & 201 & 1,228 & 897 & 0.80 & 2.12  \\
\midrule 
Piece. Const. ($n=200$) & N & 1 & 202 & 198 & 141 & 10.52 & 13.82  \\
                           & L & 1 & 401 & 1,195 & 794 & 1.86 & 2.44  \\
						   & H & 1 & 401 & 2,709 & 1,898 & 0.19 & 1.00  \\
						   
\midrule						   
Piece. Const. ($n=100$) & N & 2 & 102 & 797 & 592 & 2.61 & 3.26  \\
					   & L & 2 & 200 & 3,916 & 2,770 & 0.52 & 0.68  \\
					   & H & 2 & 200 & 4,822 & 3,473 & 0.12 & 0.37  \\
 \midrule
Coal Mining	       & N & 1 & 113 & 42 & 37 & 121.00 & 133.48  \\
	               & L & 1 & 224 & 228 & 200 & 20.07 & 24.51  \\
	               & H & 1 & 224 & 639 & 580 & 5.57 & 8.20  \\
\midrule
Tokyo Rainfall & N & 2 & 367 & 20,991 & 18,206 & 0.24 & 0.27  \\
	           & L & 2 & 731 & 53,304 & 45,629 & 0.09 & 0.11  \\
	           & H & 2 & 731 & 94,128 & 81,891 & 0.03 & 0.06  \\
    \bottomrule 

\end{longtable}
% \end{tabular}
 \end{table}

Tests indicated  that doubling the number of grid points (with a single observation per grid point) resulted in approximately 60\% fewer ESSps for each model formulation (63\% fewer for both the Normal and Laplace and 53\% for the Horseshoe), and changing from a first- to second-order model resulted in approximately 90\% fewer ESSps (91\% fewer for the Normal, 90\% for the Laplace, and 83\% for the Horseshoe).   
On average across the five scenarios investigated, the Laplace formulation resulted in 77\% fewer ESSps compared to the Normal formulation, and the Horseshoe resulted in  90\% fewer ESSps.
In terms of sampling times, the Laplace formulations on average took 4.8 times longer to achieve the same number of effective samples as the Normal formulations (range: 2.5 to 5.7 times longer), and the Horseshoe formulations took an average of 12.2 times longer than the Normal (range: 4.6 to 17.4).

\end{document}